\documentclass{amsproc}

\theoremstyle{definition}

\theoremstyle{remark}

\numberwithin{equation}{section}

\usepackage{amsfonts}
\usepackage{amsmath}
\usepackage{amssymb}
\usepackage{amscd}
\usepackage{latexsym}
\usepackage{graphicx}
\usepackage{hyperref}

\newcount\hh
\newcount\mm
\mm=\time
\hh=\time
\divide\hh by 60
\divide\mm by 60
\multiply\mm by 60
\mm=-\mm
\advance\mm by \time
\def\thistime{\number\hh:\ifnum\mm<10{}0\fi\number\mm}

\def\nn{\nonumber}

\def\Li#1(#2){\textrm{Li}_{#1}\left(#2\right)}
\def\cLi_#1(#2){\mathcal{L}_{#1}\left(#2\right)}
\def\bLi_#1(#2){\mathbf{L}_{#1}\left(#2\right)}

\def\cEs{\mathcal E_\circleddash}

\def\ZZ{{\mathbb Z}}
\def\IC{{\mathbb C}}
\def\IR{{\mathbb R}}

\def\IP{{\mathbb P}}
\def\IQ{{\mathbb Q}}

\def\cN{\mathcal{N}}
\def\cM{\mathcal{M}}
\def\cD{\mathcal{D}}
\def\cU{\mathcal{U}}

\def\cB{\mathcal{B}}

\def\cP{\mathcal{P}}

\def\cF{\mathcal{F}}

\def\Ree{\Re\textrm{e}}
\def\Imm{\Im\textrm{m}}

\def\cA{\mathcal{A}}

\def\cAtree{\mathcal{A}^{\rm tree}}

\def\fAone{\mathfrak{A}^{\rm 1-loop}}

\def\tree{\textrm{tree}}

\def\tr{\textrm{tr}}

\begin{document}

\title{AMS Proceedings Series Sample}

\title[Feynman integrals and periods]{\bf The physics and the mixed Hodge structure of 
  Feynman integrals}

\author{Pierre Vanhove}
 \address{
 Institut des Hautes Etudes Scientifiques\\
 Le Bois-Marie, 35 route de Chartres\\
 F-91440 Bures-sur-Yvette, France\hfill\break
Institut de Physique Th{\'e}orique,\\
CEA, IPhT, F-91191 Gif-sur-Yvette, France\\
CNRS, URA 2306, F-91191 Gif-sur-Yvette, France}
\email{pierre.vanhove@cea.fr}

\thanks{IPHT-t13/218, IHES/P/14/04}
\date{\today}


\dedicatory{string math 2013 proceeding contribution}


 \begin{abstract}

This expository text is an invitation to the relation between quantum
field theory Feynman integrals and periods.
We first describe the relation between the Feynman parametrization of
loop amplitudes and  world-line methods, by explaining
that the first Symanzik polynomial is the determinant of
the period matrix of the graph, and the second Symanzik polynomial
is expressed in terms of world-line Green's functions.
We then review the  relation between Feynman graphs and  variations
of mixed Hodge structures.
Finally, we provide an algorithm for generating  the Picard-Fuchs
equation satisfied by the all equal mass banana graphs in a two-dimensional space-time to all loop orders.

\end{abstract}
\maketitle
\newpage\tableofcontents\newpage

\specialsection*{{\scshape Amplitudes relations and monodromies}}

\section{{\bf Unitarity methods}}
\label{sec:unitarity-methods}

Constructions and computations of quantum field theory amplitudes have experienced
tremendous progress, leading to powerful methods for evaluating loop
amplitudes~\cite{Bern:1996je,Bern:1992ad,Bern:1994cg,Britto:2004nc}.
These methods made computable many unknown amplitudes 
and provide an increasing knowledge of gauge theory
and gravity amplitudes in various dimensions.

  These methods are  based on the unitarity properties of the scattering
amplitudes in quantum field theory.   A quantum field theory amplitude is a multivalued function presenting
branch cuts associated to particle production.

For local and Lorentz invariant quantum field theories, the matrix of
diffusion $S$  is unitary $SS^\dagger=1$. Therefore the
scattering matrix $T$, defined as $S=1+iT$ satisfies the relation
$T-T^\dagger=i TT^\dagger$.  The perturbative expansion of the
scattering matrix $T=\sum_{n\geq0} g^n A_n $ leads to unitarity
relation on the perturbative amplitudes $A_n$. This implies that the
imaginary (absorptive) part of the amplitudes $A_n$ is expressible as
some  phase integral of product of lowest order amplitudes through
Cutkosky rules~\cite{Cutkosky:1960sp}, and dispersion relation are
used to reconstruct the full amplitude. In general the evaluation of
the dispersion relations is difficult.

Fortunately,  at one-loop order, in four dimensions,  we know a basis of scalar integral functions $\{I_r\}$ specified by
boxes, triangles, bubbles, tadpoles and rational terms~\cite{Bern:1996je,Ossola:2006us,Ellis:2007qk,Ellis:2011cr}
\begin{equation}\label{e:oneloopexp}
  \fAone_n = \sum_{r}\, c_r \, I_r
\end{equation}
where $c_r$  are rational functions  of   the  kinematics   invariants.

An interesting aspect of this construction is that the scalar integral
functions have distinctive analytic properties across their branch cuts. 
For instance  the massless  four-point amplitude can  get a contribution
from the massless box $I_4(s,t)$, the one-mass triangles $I^{1m}_3(s)$
and   $I^{1m}_3(t)$,  the   massive  massive   bubbles   $I_2(s)$  and
$I_2(t)$.  The finite part of these  functions  contain contributions
with distinctive discontinuities that can be isolated by cuts
\begin{eqnarray}
I_4(s,t)&\sim& \log(-s)\log(-t)\\   
I_3^{1m}(s)&\sim&\log^2(-s)\\ 
I_2(s)&\sim&\log(-s)\,.
\end{eqnarray}
Higher-point one-loop  amplitudes have  dilogarithm functions  entering the   expression of the finite part, e.g.  $\textrm{Li}_2(1-
s_{12}s_{23}/(s_{34}  s_{56}))$. Picking a  particular  kinematic
region  $s_{12}\to\infty$,  this  function  reduces to  its  branch  cut
behaviour $\textrm{Li}_2(1-
s_{12}s_{23}/(s_{34}  s_{56}))\sim  -\log(-s_{12})\log(-s_{23})+\dots$
which can be isolated by the cut.

It is now enough to look at the discontinuities across the various
branch cuts to extract the coefficients $c_r$ in~\eqref{e:oneloopexp}.
The  ambiguity   has  to  be  a rational  function  of   the  kinematic
invariants.  There are various methods to fix this ambiguity that are discussed for
instance in~\cite{Bern:1996je}.

One of the advantages  of having a basis of integral functions is that it
permits us to state properties of the amplitudes without having to
explicitly compute them, like the no-triangle property in $\cN=8$
supergravity~\cite{Bern:2007xj,BjerrumBohr:2008vc,BjerrumBohr:2008ji,ArkaniHamed:2008gz},
or in multi-photon QED amplitudes at one-loop~\cite{Badger:2008rn}. 

\medskip

We hope that this approach can help to get a between control of the
higher-loop amplitudes contributions in field theory. At higher loop
order no basis is known for the amplitudes although it known that a
basis must exist at each loop order~\cite{Smirnov:2010hn}.

Feynman integrals from multi-loop amplitudes in quantum field theory are 
multivalued  functions. They have 
monodromy properties around the branch cuts 
in the complex energy plane, and satisfy differential equations.  This is a
strong motivation for looking at the relation between integrals from
amplitudes and  \emph{periods} of  multivalued functions.
The relation between Feynman integrals and periods is described in~\ref{sec:periods}.

\section{{\bf Monodromy and tree-level amplitude relations}}
\label{sec:mondr-ampl-relat}

Before considering higher loop integrals we start discussing
tree-level amplitudes.  Tree-level amplitudes are not periods but
they satisfy relations inherited from to the branch cuts of
the integral definition of their string theory ancestor. 
This will serve as an illustration of how the monodromy properties can constraint the
structure of quantum field theory amplitudes in Yang-Mills and gravity.

\subsection{Gauge theory amplitudes}
\label{sec:gauge-theory-ampl}

An $n$-point tree-level amplitude in (non-Abelian) gauge theory can be
decomposed into color ordered gluon amplitudes

\begin{equation}
  \mathfrak{A}^{\rm tree}_n(1,\dots,n)= g_{\rm YM}^{n-2}
  \sum_{\sigma\in\mathfrak{S}_n/\mathbb{Z}_n}                        \,
  \tr(t^{\sigma(1)}\cdots t^{\sigma(n)})\, A_{n}^{\tree}(\sigma(1,\dots,n))\,.
\end{equation}
The               color               stripped              amplitudes
$A_{n}^{\tree}(\sigma(1,\dots,n))$    are    gauge   invariant
quantities. We  are making  use of the  short hand notation  where the
entry $i$ is for the polarization $\epsilon_i$ and the momenta $k_i$,
and $\mathfrak S_n/\mathbb Z_n$ denotes the group of permutations
$\mathfrak S_n$ of $n$
letters modulo cyclic permutations. We will make use of the  notation $\sigma(a_1,\dots,a_n)$ for the
action of the permutation $\sigma$ on the $a_i$.

The color ordered amplitudes satisfy the following properties

\begin{itemize}
\item Flip Symmetry 
  \begin{equation}
    A_n^{\rm tree}(1,\dots,n)= (-1)^n \, A^{\rm tree}_n(n,\dots,1)
  \end{equation}
\item  the photon  decoupling  identity. There  is  no   
  coupling between  the Abelian field  (photon) and the  non-Abelian field
  (the gluon), therefore for $t^1=\mathbb{I}$, the identity we
  have
 \begin{equation}
    \sum_{\sigma\in\mathfrak{S}_{n-1}} \, A_n^\tree(1,\sigma(2,\dots,n))=0
\end{equation}
\end{itemize}

These relations show that the color ordered amplitudes are not
independent. The number of independent integrals  is easily determined by
representing the field theory tree amplitudes as the infinite tension limit, $\alpha'\to0$,
limit of the string amplitudes
\begin{equation}
  A_n^\tree(\sigma(1,\dots,n))= \lim_{\alpha'\to0} \cAtree(\sigma(1,\dots,n)) 
\end{equation}
where $\cAtree(\cdots)$ is the ordered string theory integral
\begin{equation}\label{e:AstringTree}
  \cAtree(\sigma(1,\dots,n)):= \int_{\Delta} \,
  f(x_1,\dots,x_n)\, \prod_{1\leq i<j\leq n-1} (x_i-x_j)^{\alpha'
    k_i\cdot k_j}\,\prod_{i=2}^{n-2} dx_i\,.
\end{equation}
In this integral we have made the following choice
for three points along the real axis $x_1=0$, $x_{n-1}=1$ and
$x_n=+\infty$  and the domain of integration is defined by 
\begin{equation}
\Delta:=\{ -\infty  <x_{ \sigma(1)}     < x_{\sigma(2)} <\cdots<x_{\sigma(n-1)}<+\infty\}\,.
\end{equation}

The function $f(x_1,\dots,x_n)$ depends only on the differences $x_i-x_j$ for
$i\neq j$. This function has poles in some of the $x_i-x_j$  but does not have any branch cut.

Since for generic values of the external momenta the scalar products
$\alpha'\,k_i\cdot k_j$ are real numbers, the factors  $(x_i-x_j)^{\alpha'
  k_i\cdot k_j}$ in the integrand require a
determination of the power $x^\alpha$ for $x<0$
\begin{equation}
x^\alpha=|x|^\alpha \begin{cases}
e^{i\pi\alpha} & {\rm Im} (x)\geq 0\,,\cr
e^{-i\pi\alpha}& {\rm Im} (x)<0\,.
\end{cases}
\end{equation}
Therefore the different orderings of the external legs, corresponding to different choices of
the permutation $\sigma$ in~\eqref{e:AstringTree}, are affected by choice of the branch cut.
The different orderings are obtained by contour deformation of
integrals~\cite{Kawai:1985xq,BjerrumBohr:2009rd,Stieberger:2009hq}. This leads to a monodromy matrix that can simply expressed
in terms of the \emph{momentum kernel} in string theory~\cite{BjerrumBohr:2010hn}

\begin{equation}
  \label{e:Sstring}
  \mathcal{S}_{\alpha'}[i_1,\ldots,i_k|
j_1,\ldots,j_k]_{p} := 
\prod_{t=1}^{k}\, {1\over\pi\alpha'}\sin\alpha'\pi\big(p\cdot
  k_{i_t}+ \sum_{q>t}^{k} \, \theta(t,q)\, k_{i_t}\cdot k_{i_q} \big)\,,
\end{equation}
and its field theory limit when $\alpha'\to0$~\cite{BjerrumBohr:2010ta,BjerrumBohr:2010zb,BjerrumBohr:2010yc}

\begin{equation}
  \label{e:SFT}
  \mathcal{S}[i_1,\ldots,i_k|
j_1,\ldots,j_k]_{p} =
\prod_{t=1}^{k}\, \big(p\cdot
  k_{i_t}+ \sum_{q>t}^{k} \, \theta(t,q)\, k_{i_t}\cdot k_{i_q} \big)\,,
\end{equation}
where $\theta(i_t,i_q)$ equals 1 if the ordering of the legs
$i_t$ and $i_q$ is opposite in the sets $\{i_1,\ldots,i_k\}$
and $\{j_1,\ldots,j_k\}$, and 0 if the ordering is the same.

As a consequence of the proprieties of the string theory integral around
the branch points one obtains that the color-ordered  amplitudes satisfy the annihilation relations both in string theory and in the field theory limit
\begin{equation}\label{e:MonSkernel}
  \sum_{\sigma\in\mathfrak                               S_{n-2}}\,\mathcal
  S[\sigma(2,\dots,n-1)|\beta(2,\dots,n-1)]_1\,
  A_n(n,\sigma(2,\dots,n-1),1)=0\,,
\end{equation}
for all  permutations $\beta\in\mathfrak S_{n-2}$.

These relations are equivalent to the BCJ relations between tree-level
amplitudes~\cite{Bern:2008qj}, and they imply that the all color-ordered amplitude can be expressed in
a basis of $(n-3)!$ amplitudes~\cite{BjerrumBohr:2009rd,Stieberger:2009hq}. 

\subsection{The gravity amplitudes}
\label{sec:gravity-amplitudes}

In the same way one can express the gravity amplitude by considering
string theory amplitudes on the sphere with $n$ marked points.

After fixing the three points $z_1=0$,
$z_{n-1}=1$ and  $z_n=\infty$, the $n$-point closed string
amplitude takes the general form
\begin{equation}
  \label{eq:DefMn}
  \mathcal{M}_n=\left(i\over 2\pi\alpha'\right)^{n-3}\,
\int \prod_{1\leq i<j\leq    n-1}\!\! |z_j-z_i|^{2\alpha'\,k_i\cdot k_j}\, f(z_i)\,g(\bar z_i)\, \prod_{i=2}^{n-2} d^2z_i\, ,
\end{equation}
where $f(z_i)$ and $g(\bar  z_i)$ arise from the operator
product expansion of the vertex operators. They are functions
without branch cuts of the differences $z_i-z_j$ and $\bar z_i-\bar
z_j$ with possible poles in these variables. The precise form of these functions
depends on the external states. 

\begin{figure}[t]
\centering
\includegraphics[width=12cm]{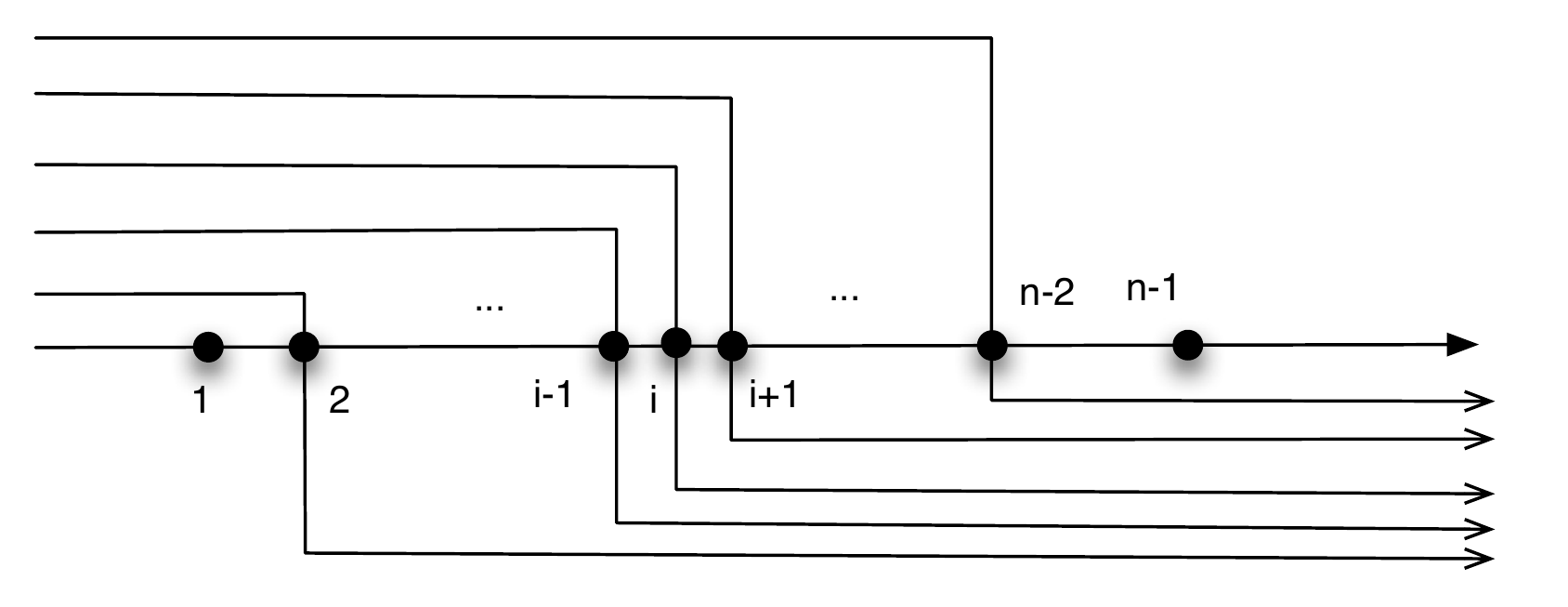}
\caption{\sl The nested structure of the contours of integration
for the variable $v^-_i$   corresponding       to        the       ordering
$0<v^+_2<v^+_3<\cdots<v^+_{n-2}<1$ of the $v_+$ variables.}
\label{fig:contourNested}
\end{figure}

Changing variables to $z_i = v_i^1 +iv_i^2$, one can factorize the
integral (we refer to~\cite{BjerrumBohr:2010hn} for details) 
\begin{align}
\label{eq:DefMn1}
&\mathcal{M}_n= \left(-1\over 4\pi\alpha'\right)^{n-3}\!\!
\int_{-\infty}^{+\infty} \prod_{i=2}^{n-2} dv_i^+dv_i^-
f(v^-_i)\,g(v_i^+)\nonumber \\
&\hspace{1cm}\times 
(v_i^+)^{\alpha' k_1\cdot k_i}
(v_i^-)^{\alpha' k_1\cdot k_i}
(v_i^+-1)^{\alpha'\,k_{n-1}\cdot k_i}
(v_i^--1)^{\alpha'\,k_{n-1}\cdot k_i}\nonumber \\
&\hspace{1cm}\times \prod_{i<j\leq n-2} \big(v_i^+-v_j^+
\big)^{\alpha'\,k_i\cdot k_j}
\big(v_i^--v_j^-\big)^{\alpha'\,k_i\cdot k_j}\,.
\end{align}
We now consider the deformations of the contours of integration
for the $v^-_i$ variables given in figure~\ref{fig:contourNested}.  Because the contours cannot cross
each other we need to close them either to the right, turning
around the branch cut at $z=1$ by starting with the rightmost,
or close the contours to the left, turning around the branch
cut at $z=0$, starting with the leftmost.

There is evidently an arbitrariness in the number of contours
that are closed to the left or to the right. For a given
$2\leq j\leq n-2$, we can pull the contours for the set between
2 and $j-1$ to the left, and the set between $j$ and $n-2$ to
the right. The  independence of the amplitude under this choice  is a consequence of the monodromy
relations in eq.~\eqref{e:MonSkernel}.

To get  the full closed  string amplitude~(\ref{eq:DefMn}) we
need to multiply the left-moving amplitude of the $v^+$
integrations with the right-moving
contribution from the integration over $v^-$ and then sum over all
orderings to get
\begin{multline}
  \label{eq:Mnfinal}
 \mathcal{M}_n ={} \left(-i/4\right)^{n-3}\times
\sum_{\sigma\in\mathfrak S_{n-3}}\sum_{\gamma\in\mathfrak S_{j-2}}
\sum_{\beta\in\mathfrak S_{n-j-1}}\cr
\mathcal{S}_{\alpha'}[\gamma\circ\sigma(2,\dots,j\!-\!1)|\sigma(2,\dots,j\!-\!1)]_{k_1}
\mathcal{S}_{\alpha'}[\beta\circ\sigma(j,\dots,n\!-\! 2)|
\sigma(j,\dots,n\!-\! 2)]_{k_{n\!-\!1}}\cr
\times
   \cA_n(1,\sigma(2,\dots,n-2),n- 1,n)\,
   \widetilde{\cA}_n(\gamma\circ\sigma(2,\dots,j- 1),1,n-1,\beta\circ\sigma(j,\dots,n-2),n)\,.
\end{multline}
where the amplitudes $\cA(\cdots)$ (respectively $\widetilde \cA(\cdots)$) are obtained from integration
over the variables $v_i^+$ (respectively $v_i^-$).
This provides a general form of the closed/open string
relation between external gauge bosons and gravitons at
tree-level. When restricted to the case of graviton external
states the field theory limit of this expression reduces to the
form derived in~\cite{BjerrumBohr:2010zb,BjerrumBohr:2010yc}.

The choice of contour deformation made by KLT in~\cite{Kawai:1985xq} consists in closing
half of the contours to the left and and the other
half to the right. This leads to the smallest number of
terms in the sum~\eqref{eq:Mnfinal}.

For $j=n-1$ the field theory gravity amplitude takes a form
characteristic of the expression of  gravity amplitudes as  a sum of
square of Yang-Mills amplitudes
\begin{multline}
\label{stringPureKLTn}
{M}_n =(-1)^{n-3} \sum_{\sigma,\gamma\in\mathfrak S_{n-3}}
\mathcal{S}[\gamma(2,\dots,n-2)|\sigma(2,\dots,n-2)]_{k_1}\cr
\times A_n(1,\sigma(2,\dots,n-2),n-1,n)
\widetilde{A}_n(n-1,n,\gamma(2,\dots,n-2),1)\,.
\end{multline}
The previous construction provided amplitudes relations between massless tree-level amplitudes
in gauge and gravity amplitudes.

This construction does not make any explicit reference to a given
space-time dimension, seeing a massive particle in, say in dimension $D=4$, as the dimensional
reduction of a massless particle in higher dimensions, it is immediate 
 that the amplitude relations formulated with  the momentum kernel are valid
for massive external particles. This has been  applied to amplitudes
between massive matter field in pure gravity~\cite{Bjerrum-Bohr:2013bxa}.

This construction made an important use of the multivalueness of the
factors $\prod_{1\leq i<j\leq n} (x_i-x_j)^{a_{ij}}$  with
$a_{ij}=\alpha\, k_i\cdot k_j\in\IR$. Although the tree-level integrals are
not periods, their infinity tension expansion for $\alpha'\to0$ is
expressible in terms of multiple zeta values that are periods~\cite{BrownSMZV,Broedel:2013tta,StiebergerSMZV,TaylorStiebergerSZM}.

\specialsection*{{\scshape Feynman integrals and periods}}
\section{{\bf Feynman integral}}
\label{sec:feynman-integral}

\subsection{The Feynman parametrization}
\label{sec:feynm-param}

A connected Feynman graph $\Gamma$ is determined by the number  $n$ of propagators
(internal edges), 
the number $l$ of  loops, and the number $v$ of
vertices. The Euler characteristic of the graph relates these three
numbers as $l=n-v+1$, therefore only the number of loops $l$ and the
number $n$ of propagators are needed.

In a momentum representation an $l$-loop with $n$ propagators Feynman
graph reads\footnote{In this text we will consider only graph without
  numerator factors. A similar discussion can be extended to this case
  but will not be considered here.}
\begin{equation}\label{e:GraphFeyn}
  I^D_\Gamma(p_i,m_i):=
 { (\mu^2)^{\sum_{i=1}^n\nu_i-l{D\over2}}\over \pi^{lD\over2}}\,{\prod_{i=1}^n
    \Gamma(\nu_i)\over \Gamma(\sum_{i=1}^n\nu_i-l{D\over2})} \,
  \int_{(\IR^{1,D-1})^l} \,   {\prod_{i=1}^l  d^D\ell_i\over
    \prod_{i=1}^n (q_i^2-m_i^2+i\varepsilon)^{\nu_i}}
\end{equation}
where $\mu^2$ is a scale of dimension mass squared. 
Some of the vertices are connected to external momenta $p_i$ with
$i=1,\dots,v_e$ with $0\leq v_e\leq v$. The internal masses  are positive 
 $m_i\geq0$ with $1\leq i\leq n$. Finally
 $+i\varepsilon$ with $\varepsilon>0$ is the
Feynman prescription for the propagators for a space-time metric of
signature $(+-\cdots-)$, and $D$ is the space-time dimension, and we set $\nu:=\sum_{i=1}^n \nu_i$.

\medskip 
Introducing the  size $l$ vector of loop momenta $L^\mu:=(\ell_1^\mu,\dots,\ell_l^\mu)^T$
corresponding to the minimal set of linearly independent
momenta flowing along the graph.  We introduce as well the size $v_e$
vector of external momenta $P^\mu=(p_1^\mu,\dots,p_{v_e}^\mu)^T$. Since we take the convention that all
momenta are incoming  momentum conservation implies that $\sum_{i=1}^{v_e} p_i=0$.

Putting the  momenta $q_i$ flowing along the graph in  a  size $n$
vector $q^\mu:=(q_1^\mu,\dots,q_n^\mu)^T$. Momentum conservation at
each vertices of the graph gives the relation

\begin{equation}\label{e:qiDef}
  q^\mu= \rho \cdot L^\mu + \sigma\cdot P^\mu\,.
\end{equation}
 The matrix $\rho$ of size $n\times l$ has entries taking values in $\{-1,0,1\}$, the
signs depend on an orientation of the propagators. (The orientation of the graph and the
choice of basis for the loop momenta
will be discussed further in section~\ref{sec:word-line-formalism}.)
The matrix $\sigma$ of size $n\times v_e$ has only entries taking values
in $\{0,1\}$ because have the convention that all external momenta are incoming.

\medskip
We introduce the Schwinger proper-times $\alpha_i$  conjugated to each
internal propagators

\begin{equation}
    I^D_\Gamma(p_i,m_i)={(\mu^2)^{\nu-l{D\over2}}\over\pi^{lD\over2}\Gamma(\nu-l{D\over2})}\,  \int_{(\IR^{1,D-1})^l} \, \int_{[0,+\infty[^n}     \!\!\!\!\!
  e^{-\sum_{i=1}^n \alpha_i (q_i^2-m_i^2+i\varepsilon)} \prod_{i=1}^n
  {d\alpha_i \over \alpha_i^{1-\nu_i}}\prod_{i=1}^l  d^D\ell_i\,.
\end{equation}
Setting $T=\sum_{i=1}^n \alpha_i$ and $\alpha_i=T\,x_i$ this integral
becomes
\begin{equation}
    I^D_\Gamma(p_i,m_i)={(\mu^2)^{\nu-l{D\over2}}\over\pi^{lD\over2}\Gamma(\nu-l{D\over2})}\,  \int_{(\IR^{1,D-1})^l} \, \int_{[0,+\infty[^{n+1}}     \!\!\!\!\!
  e^{-T\mathcal Q} \delta(\sum_{i=1}^n x_i-1) \, {dT\over T^{1-\nu}}\prod_{i=1}^n
   {dx_i\over  x_i^{1-\nu_i}} \prod_{i=1}^l  d^D\ell_i\,,
\end{equation}
where we have defined
\begin{equation}\label{e:Qdef}
  \mathcal Q:=  \sum_{i=1}^n x_i (q_i^2-m_i^2)\,.
\end{equation}
Introducing the $n\times n$ diagonal matrix $X=\textrm{diag}(x_1,\cdots,
x_n)$, one rewrites this  expression exhibiting the quadratic form in the loop momenta
\begin{equation}\label{e:QOmega}
\mathcal Q= (L^\mu+\Omega^{-1} Q^\mu)^T\cdot \Omega\cdot (L^\mu+\Omega^{-1}Q^\mu)- J-
(Q^\mu)^T\cdot \Omega^{-1}\cdot Q^\mu\,,
  \end{equation}
where   we have defined
\begin{equation}\label{e:DefOmega}
\Omega:= \rho^TX\rho,\quad Q^\mu:=
\rho^TX\sigma\,P^\mu, \quad 
J:=(P^\mu)^T \sigma^TX\sigma P^\mu+\sum_{i=1}^n x_i (m_i^2-i\varepsilon)
\end{equation}
 and 
we made use of the fact that the square $l\times l$ matrix $\Omega$ is symmetric and
invertible. 
Performing the Gaussian integral over the loop momenta $L^\mu$ one gets
\begin{equation}
    I^D_\Gamma(p_i,m_i)={(\mu^2)^{\nu-l{D\over2}}\over\Gamma(\nu-l{D\over2})}\, \int_{[0,+\infty[^{n+1}}     \!\!\!\!\!
  e^{-T \mu^2\,\cF \, \cU^{-1}} {\delta(\sum_{i=1}^n x_i-1)\over
\cU^{l{D\over2}}}\prod_{i=1}^n
  {dx_i\over  x_i^{1-\nu_i}}\, {dT\over T^{1-\nu+l{D\over2}}} \,.
\end{equation}
Introducing the notations for the first Symanzik polynomial
\begin{equation}\label{e:defOmega}
  \cU:=\det(\Omega)  
\end{equation}
and using the  adjugate matrix of
$\textrm{Adj}(\Omega):= \det \Omega\, \Omega^{-1}$, we define the
second Symanzik polynomial
\begin{equation}\label{e:defF}
  \cF:=  {-J\,\cU+ (Q^\mu)^T\cdot  \textrm{Adj}(\Omega)\cdot Q^{\mu}\over \mu^2}\,.
\end{equation}
A modern approach to the derivation of these polynomials using graph
theory is given in~\cite{Bogner:2010kv}.
 In section~\ref{sec:word-line-formalism} we will give an
interpretation of these quantities using the first quantized world-line formalism. 

Performing the integration over $T$, one arrives at the
expression for a Feynman graph given in quantum
field theory textbooks like~\cite{Itzykson:1980rh}

\begin{equation}\label{e:GraphIZ1}
  I^D_\Gamma(p_i,m_i)=\int_{[0,+\infty[^n}   \,
  {\cU^{\nu- (l+1){D\over2}}\over \cF^{\nu-l
      {D\over2}}}\,\delta(\sum_{i=1}^n x_i-1)\prod_{i=1}^n x_i^{\nu_i-1}  dx_i\,.  
\end{equation}

\begin{itemize}
\item 
Notice that $\cU$ and $\cF$ are independent of the dimension of
space-time. The space-time dimension enters only in the powers of
$\cU$ and $\cF$ in the expression for the Feynman graph in~\eqref{e:GraphIZ1}.

\item
The graph polynomial $\cU$ is an homogeneous polynomial of degree $l$
in the Feynman parameters $x_i$. $\cU$ is linear in each of the $x_i$. This graph polynomial does not depend
on the internal masses $m_i$ or the external momenta $p_i$. In
section~\ref{sec:word-line-formalism}, we argue that this polynomial
is the determinant of the period matrix of the graph.
\item
The graph polynomial $\cF$ is of degree $l+1$. This polynomial depends
on the internal masses $m_i$ and the kinematic invariants $p_i\cdot
p_j$. If all
internal masses are vanishing then $\cF$ is linear in the Feynman
parameters $x_i$  as  is $\cU$.
\end{itemize}
Since  the coordinate scaling $(x_1,\dots,x_n)\to \lambda
(x_1,\cdots,x_n)$ leaves invariant the integrand and the domain of integration, we can rewrite this
integral as

\begin{equation}\label{e:GraphIZ2}
  I^D_\Gamma(p_i,m_i)=\int_{\Delta}   \prod_{i=1}^n x_i^{\nu_i-1} \,
  {\cU^{\nu- (l+1){D\over2}}\over \cF^{\nu-l {D\over2}}}\, \omega  
\end{equation}
where $\omega$ is the differential $n-1$-form

\begin{equation}\label{e:diffInt}
  \omega:= \sum_{j=1}^n  (-1)^{j-1} \, x_j\, dx_1\wedge \cdots \wedge
  \widehat {dx_j}\wedge\cdots \wedge dx_n  
\end{equation}
where $\widehat{dx_j}$ means that $dx_j$ is omitting in this sum.
The domain of integration $\cD$ is defined as
\begin{equation}\label{e:DefDomain}
  \Delta:=\{[x_1,\cdots,x_n]\in \mathbb P^{n-1}| x_i\in\mathbb R,  x_i\geq0\}.  
\end{equation}
%

\subsection{Ultraviolet and infrared divergences}
\label{sec:diverg-epsil-expans}

The Feynman integrals in~(\ref{e:GraphFeyn})
and~(\ref{e:GraphIZ1}) evaluated in four dimensions $D=4$ have in general ultraviolet and infrared
divergences. One can work in a space-time dimension around four
dimension by setting $D=4-2\epsilon$ in the
expression~(\ref{e:GraphIZ1}) and performing a Laurent series
expansion around $\epsilon=0$
\begin{equation}
  I_\Gamma^{(4-2\epsilon)} (p_i,m_i) = \sum_{k\geq -2l} \epsilon^{k} \, I_\Gamma^{(k)}(p_i,m_i)+O(\epsilon)\,.
\end{equation}
At one-loop around four dimensions the structure 
of the integrals is now very 
well understood~\cite{Bern:1997sc,Britto:2004nc,Ossola:2006us,Bern:1996je,Britto:2010xq,Ellis:2011cr}. General formulas for all one-loop amplitudes can be
found in~\cite{Ellis:2007qk,QCDsite}, some higher-loop recent
considerations can be found in~\cite{Panzer:2014gra}.

\medskip

In this work we discuss properties of the Feynman integrals valid for
any values of $D$ but when we  evaluate Feynman integrals in
sections~\ref{sec:schw-repr}---\ref{sec:explicit-banana} we will work
with both ultraviolet and infrared finite integrals.  If all the
internal masses are positive $m_i>0$ for $1\leq i\leq n$ then the integrals
are free of infrared divergences.  Working in $D=2$ will make the
integrals free of ultraviolet divergences. One can then relate the
expansion around four dimensions to the one around two dimensions
using the dimension shifting relations~\cite{Tarasov:1996br}.

\subsection{The word-line formalism}
\label{sec:word-line-formalism}

The world-line formalism is a first quantized approach to
amplitude computations. This formalism has the advantage of being close
in spirit to the one followed by string theory perturbation.

Some of the rules for the world-line formalism can be deduced from a
field theory limit infinite tension limit, $\alpha'\to0$, of string theory. At
one-loop order this leads to the so-called `string based rules' used
to compute amplitude in
QCD~\cite{Bern:1987tw,Bern:1990ux,Bern:1991aq,Bern:1991an,Strassler:1992zr}
and in gravity~\cite{Bern:1993wt,Dunbar:1994bn}, as reviewed
in~\cite{Bern:1992ad,Schubert:2001he}.  One can motivate this
construction by taking a field theory limit of
string theory amplitudes as for instance
in~\cite{Frizzo:1999zx,DiVecchia:1996dp,Magnea:2013lna,Tourkine:2013rda}.

At one-loop order, this formalism has the advantage of making obvious
some generic properties of the amplitudes like the no-triangle
properties in $\cN=8$
supergravity~\cite{BjerrumBohr:2008vc,BjerrumBohr:2008ji} or for
multi-photon amplitudes in QEDa at one-loop~\cite{Badger:2008rn}.  These rules have been extended to
higher-loop orders
in~\cite{Schmidt:1994zj,Roland:1996np,Roland:1997wq,Sato:1998sf}.  See
for instance~\cite{Green:1999pu,Green:2008bf} for a treatment of the
two-loop four-graviton $\cN=8$ supergravity amplitude in various
dimensions.  This formalism is compatible with the pure spinor
formalism providing a first quantized approach for the
super-particle~\cite{Berkovits:2001rb,Bedoya:2009np} that can be
applied to amplitude computations in maximal
supergravity~\cite{Anguelova:2004pg,Bjornsson:2010wm,Bjornsson:2010wu,Cederwall:2012es}.

\bigskip
In this section we will follow a more direct approach to describe the
relation between Feynman graphs and the world-line approach. A more
systematic derivation will be given elsewhere.

First consider an $l$-loop vacuum graph $\Gamma_0$ without external momenta with
$n_0$ propagators (edges) and $v_0=n_0-l+1$ vertices. One needs to assign a labeling and an orientation of the vacuum graph
corresponding to a choice of $l$ independent loop momenta circulating
along the loop, this
orientation conditions the signs in the incidence matrix
in~\eqref{e:qiDef}. Label the Schwinger proper-time of each
propagator by $T_i$ with $i=1,\dots,n_0$. Therefore the expression
for $\mathcal Q$ in~\eqref{e:Qdef} becomes
\begin{equation}\label{e:Qvac}
  \mathcal Q_0= (L^\mu)^T\cdot \Omega\cdot (L^\mu)- \sum_{ i=1}^{n_0} T_i (m_i^2-i\varepsilon)\,.
\end{equation}
where $\Omega= \rho^T \textrm{diag}(T_1,\dots, T_{n_0})\rho$ is 
the \emph{period matrix} associated with the graph.
The presentation will follow the one
given in~\cite[section~2.1]{Green:2008bf} for two-loop graphs.
Let choose a basis of oriented closed  loops $C_i$ with
$1\leq i\leq l$ for the graph. In this context the
loop number $l$ is the first Betti number of the graph. 
And  let $\omega_i$ be the elementary line
element along the closed loop $C_i$. 
The entries of the matrix $\Omega$ constructed above are given by the oriented
circulation of these line elements along each loop $C_i$ 
\begin{equation}
\Omega_{ij}=\oint_{C_i} \omega_j\,.
\end{equation}
A direct  construction of this matrix from graph theory is detailed in~\cite[section~IV]{Dai:2006vj}.

We now consider Feynman graphs with external momenta.  One can construct 
such  graphs by starting from a particular vacuum graph $\Gamma_0$ and adding to
it external momenta.
\begin{itemize}
\item 
One can  add external momenta to some
vertices of the vacuum graph. This operation does not modify the
numbers of vertices and  propagators of the graph. This will affect
external momentum dependence part in the definition of the existing momenta
$q_i$ in~\eqref{e:Qvac}. This operation does not modify definition of
the period matrix $\Omega$.
\item One can consider adding  new vertices with incoming
  momenta. One vertex attached to external momenta,  has to be added
  on a given internal edge (propagator) say $i_*$. Under this operator
  the number of vertices has increased by one unit as well as the
  number of propagators. The number of loops has not been modified. 
This operation splits the internal propagator $i_*$ into two as depicted
\begin{center}
  \includegraphics[width=10cm]{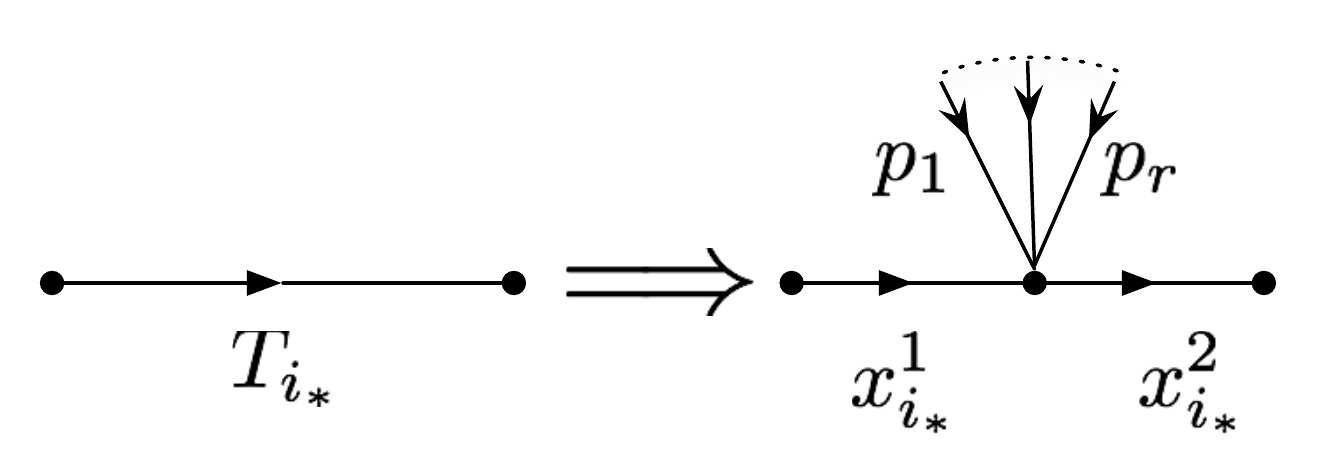}
\end{center}
Under this operation the proper-time $T_{i_*}=x^1_{i_*}+x^2_{i_*}$ and
the momentum flowing along the edge $i_*$ is replaced by $q_{i_*}^1=
q_{i_*}$ and $q_{i_*}^2= q^1_{i_*} +P$ where $P= \sum_{j=1}^r P_j$ is
the sum of all incoming momenta on the added vertex.  The insertion
point of the new vertex is parametrized by $x^2_{i_*}$.
 The
expression for $\mathcal Q$ in~\eqref{e:Qdef} then becomes
\begin{equation}
  \mathcal Q=\sum_{i=1\atop i\neq i_*}^n T_i (q_i^2-m_i^2+i\varepsilon)  +
  x_{i_*}^1 ((q^1_{i_*})^2-m_{i_*}^2+i\varepsilon)+  x_{i_*}^2
  ((q^2_{i_*})^2-m_{i_*}^2+i\varepsilon)\,.
\end{equation}
Since $  x_{i_*}^1 ((q^1_{i_*})^2-m_{i_*}^2+i\varepsilon)+  x_{i_*}^2
  ((q^2_{i_*})^2-m_{i_*}^2+i\varepsilon)= T_{i_*}
  ((q_{i_*})^2-m_{i_*}+i\varepsilon) + 2 x^2_{i_*} q_{i_*}\cdot P$, we conclude that the matrix $\Omega$ entering
  the expressions for $\mathcal Q$ in~\eqref{e:QOmega} for a graph
  with external momenta is the same as the one of the associated
  vacuum graph.
\end{itemize}

We therefore conclude that in the representation of a Feynman graph $\Gamma$
in~\eqref{e:GraphIZ1} and~\eqref{e:GraphIZ2} the first Symanzik
polynomial $\cU$ is the determinant of the period matrix of the vacuum
graph $\Gamma_0$ associated to the graph $\Gamma$.

\medskip
We now turn to the reinterpretation of the  second Symanzik polynomial $\cF$ 
in~\eqref{e:defF}  using the world-line methods.

Define $\hat\cF:=\cF/ \cU$  one can rewrite this
expression in the following way
\begin{equation}
\hat\cF  =-\sum_{i=1}^n x_i  (m_i^2-i\varepsilon)+ \sum_{1\leq r,s \leq m} p_r\cdot p_s \, G(x_r,x_s;\Omega)  \,.
\end{equation}
where  we have introduced the \emph{Green function} 
\begin{equation}\label{e:Gdef}
  G(x_r,x_s;\Omega)= -\frac12 d(x_r,x_s)  +\frac12\left(\int_{x_s}^{x_r}\omega\right)\cdot\Omega^{-1}\cdot\left(
  \int_{x_s}^{x_r} \omega\right)\,,
\end{equation}
where $\omega= (\omega_1,\dots,\omega_l)^T$ the size $l$ vectors of
elementary line elements along the loops, and $d(x_r,x_s)$ is the 
distance between the two vertices of coordinates $x_r$ and $x_s$ on
the graph.  
This is the expression for the  Green function between  two points on the word-line graph 
constructed  in~\cite[section~IV]{Dai:2006vj}.
In section~\ref{sec:example-green-funct} we provide a few examples.

\medskip
One can therefore give an alternative form for the parametric
representation of the Feynman graph $\Gamma$, by writing splitting the
integration over the parameters into an integration over the
proper-times $T_i$ with $1\leq i\leq n_0$ of the vacuum graph, and the
insertion points $x_i$ with $1\leq i\leq m$ of the extra vertices
carrying external momenta
\begin{equation}
     I^D_\Gamma(p_i,m_i)=\int_{\cD}  
  {1\over \hat\cF^{3(l-1)+m-l {D\over2}}}\,\prod_{i=1}^m dx_i\, {\prod_{i=1}^{n_0} dT_i\over
    (\det \Omega)^{D\over2}}\,.
\end{equation}
For the case of $\varphi^3$ vertices, at the loop order $l$,  the number of propagators of
vacuum graph is $n_0=3(l-1)$ and this formulation leads to a treatment of field theory graphs in a string theory manner
as in~\cite{BjerrumBohr:2008vc,BjerrumBohr:2008ji,Badger:2008rn,Bjornsson:2010wm,Bjornsson:2010wu}.

\subsubsection{Examples of period matrices}
\label{sec:examplesperiodmatrices}

The construction applies to any kind of interaction since we never
used the details of the valence of the vertices one needs to consider.
We provide a few example based on $\varphi^3$ and $\varphi^4$ scalar theories.

\begin{figure}[ht]
  \centering
  \includegraphics[width=8cm]{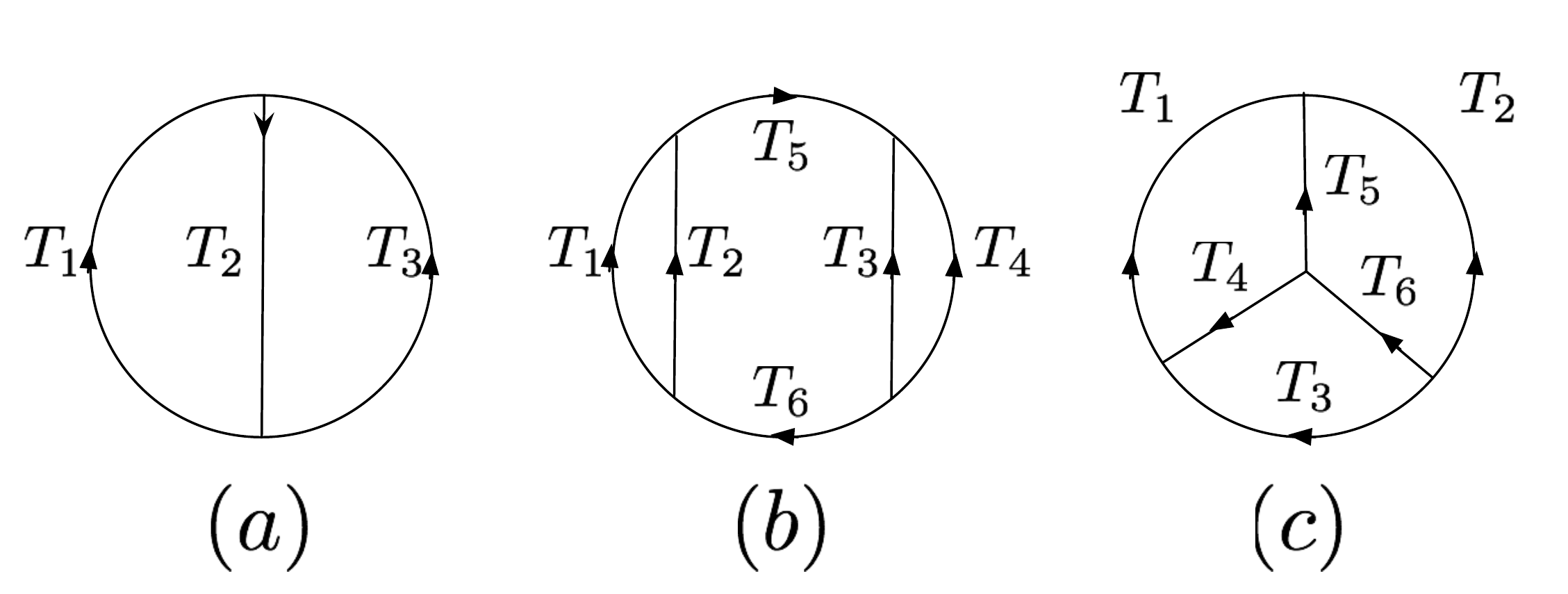}
  \caption{Examples of $\varphi^3$ vacuum graphs at (a) two-loop order, (b) and (c) at
    three-loop order.}
  \label{fig:vacgraph}
\end{figure}

For instance for the two-loop and three-loop graphs of figure~\ref{fig:vacgraph}
the period matrix are given by  

\begin{eqnarray}\label{e:Omega2}
 \hskip.5cm \Omega_2&=&
  \begin{pmatrix}
    T_1+T_3& T_3\cr T_3 & T_2+T_3
  \end{pmatrix}~\textrm{in~figure}~\ref{fig:vacgraph}(a)\\
\nn  \Omega_{3} &=&
  \begin{pmatrix}
    T_1+T_2& T_2&0\cr T_2&T_2+T_3+T_5+T_6 & T_3\cr
0&T_3&T_3+T_4
  \end{pmatrix}~\textrm{in~figure}~\ref{fig:vacgraph}(b)\cr
 \Omega_{  3} &=&
  \begin{pmatrix}
    T_1+T_4+T_5& T_5&T_4\cr T_5&T_2+T_5+T_6& T_6\cr T_4&T_6&T_3+T_4+T_6
  \end{pmatrix}~\textrm{in~figure}~\ref{fig:vacgraph}(c)\,.
\end{eqnarray}
A list of period matrices for $\varphi^3$ vacuum graphs up to and
including four loops can be found in~\cite{Bjornsson:2010wm,Bjornsson:2010wu}.

\begin{figure}[ht]
  \centering
  \includegraphics[width=10cm]{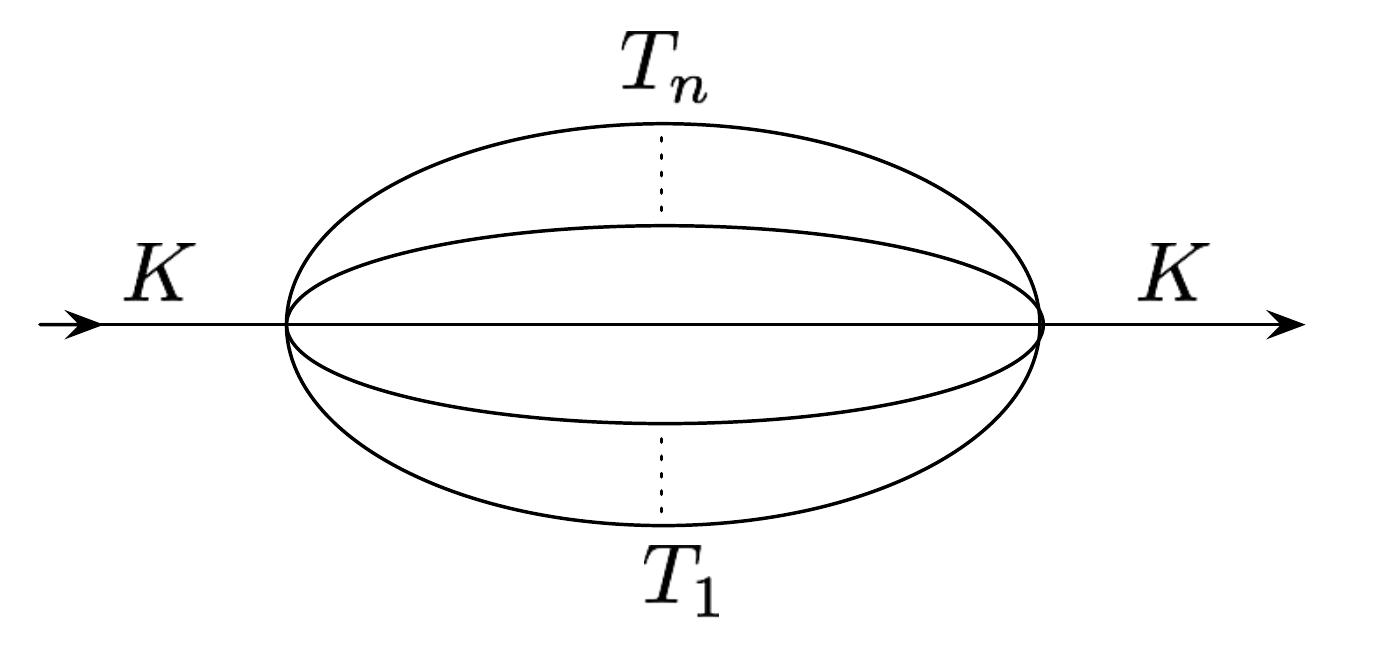}
  \caption{Graph for the banana graph with $n$ propagators.}
  \label{fig:banana}
\end{figure}

For the banana graphs with $n$ propagators (and $n-1$ loops) in figure~\ref{fig:banana} the period
matrix is given  by
\begin{equation}\label{e:OmegaBanana}
  \Omega_{banana}=
  \begin{pmatrix}
    T_1+T_n & T_n&\cdots &&T_n\cr
    T_n & T_2+T_n& T_n&\cdots &T_n\cr
\vdots & &&&\vdots\cr
T_n&\cdots&&T_n&T_{n-1}+T_n
  \end{pmatrix}\,.
\end{equation}
These graphs will be discussed in detail in  section~\ref{sec:banana}.

\subsubsection{Example of Green functions}
\label{sec:example-green-funct}

\begin{figure}[ht]
  \centering
  \includegraphics[width=10cm]{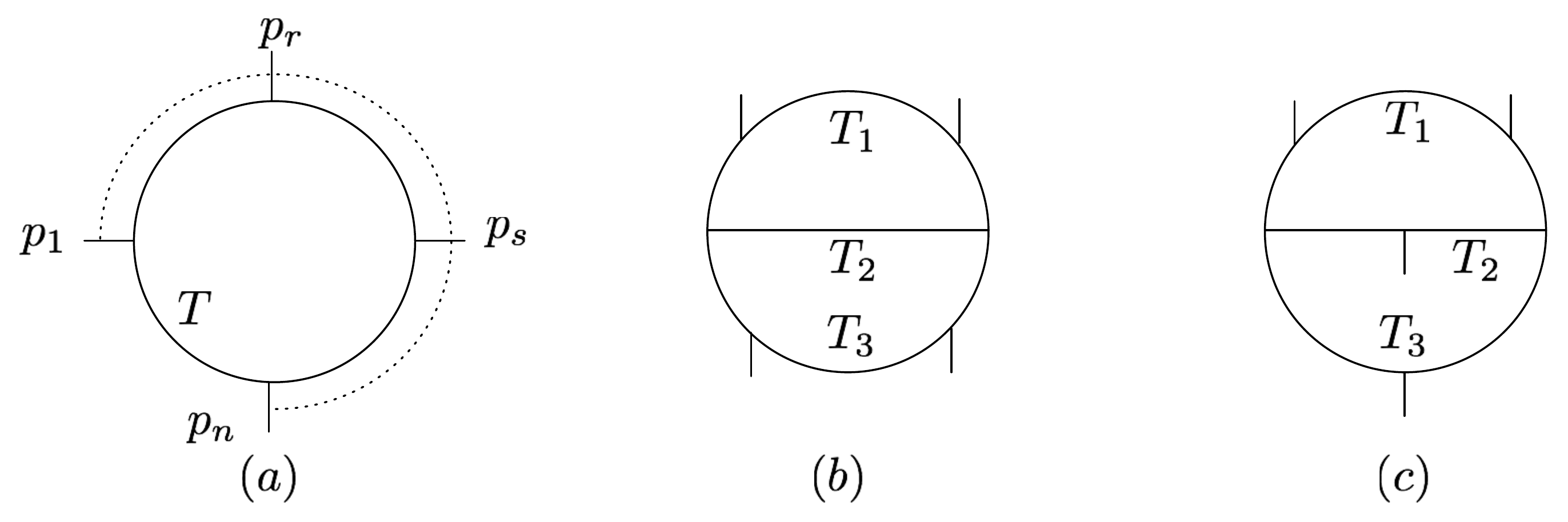}
  \caption{(a) One-loop $n$-point graph and (b)-(c) two-loop  four-point graphs.}
  \label{fig:graphs}
\end{figure}

We provide a example of the construction of the second Symanzik
polynomial $\hat\cF$ using the Green function method.

For the one-loop graph of figure~\ref{fig:graphs}(a) the period matrix
$\Omega= T$ is the length of the loop. The Green function between the
external states with momenta $p_r$ and $p_s$ is given by 
\begin{equation}
  G^{\rm 1-loop}(x_r,x_s;L)= -\frac12 \, |x_s-x_r|+ \frac12
  \,{(x_r-x_s)^2\over T}\,.
\end{equation}
For massless external states $p_r^2=0$ for $1\leq r\leq n$, the reduced second Symanzik polynomial $\hat\cF^{1-loop}$ is given by 
\begin{equation}
  \hat \cF^{\rm 1-loop}=  \sum_{1\leq r<s\leq n} p_r\cdot p_s\, G^{1-loop}(x_r,x_s;L)\,.
\end{equation}

In the massless case the two-loop Green's function have been
derived in~\cite[section~2.1]{Green:2008bf}. If the two external
states with momenta $p_r$ and $p_s$ are on the same line, say the one
of length $T_1$, then 
\begin{equation}
  G^{\rm 2-loop}(x_r,x_s;\Omega_2)=-\frac12  |x_s-x_r| +{T_2+T_3\over 2}{(x_s-x_r)^2\over (T_1T_2+T_1T_3+T_2T_3)}\,,
\end{equation}
where we used the two-loop period matrix $\Omega_2$
in~\eqref{e:Omega2} such that $\det\Omega_2= T_1T_2+T_1T_3+T_2T_3$. 
If the external states with momenta $p_r$ and $p_s$ are on
different  lines, say  $p_r$ is on $T_1$ and $p_s$ on $T_2$, then the
Green's function is given by 
\begin{equation}
  G^{\rm 2-loop}(  x_r,x_s;\Omega_2)= -\frac12(x_r+x_s)+ {T_3(x_r+x_s)^2+
    T_2 x_r^2+T_1 x_s^2\over 2(T_1T_2+T_1T_3+T_2T_3)}\,.
\end{equation}
With these Green functions the reduced second Symanzik polynomial for
the four-point two-loop graphs in figure~\ref{fig:graphs}(b)-(c) read
\begin{equation}
  \hat\cF^{2-loop}= \sum_{1\leq r<s\leq 4} p_r\cdot p_s \, G^{2-loop}(x_r,x_s;\Omega_2)\,.
\end{equation}
\section{{\bf Periods}}
\label{sec:periods}

In the survey~\cite{Kontsevich:2001}, Kontsevich and Zagier give the following
definition of  the ring $\cP$ of periods: \emph{a period is a complex number that can be expressed as an
integral of an algebraic function over an algebraic domain.}

In more precise terms $z\in\cP$ is a period if its real part $\Ree(z)$
and imaginary part $\Imm(z)$ are of the form
\begin{equation}
  \label{e:PeriodKZ}
\int_{\Delta} {f(x_1,\dots,x_n)\over g(x_1,\dots,x_n)} \,
\prod_{i=1}^n dx_i\,,  
\end{equation}
where $f(x_1,\dots,x_n)$ and $g(x_1,\dots,x_n)$ belong to $\mathbb
Z[x_1,\cdots,x_n]$ and the domain of integration $\Delta$ is a domain
in $\IR^n$ given by polynomial inequalities with rational coefficients.

 Since sums and
products of periods remain periods, therefore the periods form a
ring, and the periods form a sub $\bar {\mathbb Q}$-algebra of
$\mathbb C$ (where $\bar{\mathbb Q}$ is the set of algebraic numbers).

Examples of  periods represented by single integral
\begin{equation}
  \sqrt 2=\int_{2x^2\leq1} dx; \quad \log(2)= \int_{1\leq x\leq 2} {dx\over x}  
\end{equation}
or by a double integral
\begin{equation}
  \zeta(2) ={\pi^2\over6}= \int_{0\leq t_1\leq t_2\leq 1}  \,   {dt_2\over t_2} \, {dt_1\over 1-t_1}\,.
\end{equation}
This example the value at 1 of the dilogarithm $\Li2(1)=\zeta(2)$
where
\begin{equation}
\Li2(x) = \sum_{n \geq1}  {x^n\over n^2}; \qquad \textrm{for}\qquad
0\leq x<1\,.
\end{equation}

In particular, it is familiar to the quantum field theory practitioner that
the finite part of one-loop amplitudes in four dimensions
is expressed in terms of  dilogarithms.

Under change of variables and integration  a period can take a form
given in~(\ref{e:PeriodKZ}) or not. One example is $\pi$ which can be
represented by the following two-dimensional or one-dimensional integrals
\begin{equation}
  \pi = \int_{x^2+y^2\leq 1} dxdy = 2\,\int_0^{+\infty} {dx\over 1+x^2}  = \int_{-1}^1 {dx\over\sqrt{1-x^2}}
\end{equation}
or by the following contour integral
\begin{equation}
  2i\pi = \oint {dz\over z}\,.  
\end{equation}

At a first sight, the definition of a period given
in~(\ref{e:PeriodKZ}) and the Feynman representation of the Feynman
graph in~(\ref{e:GraphIZ1}) \emph{look similar}. A relation between these two
objects  was remarked in the pioneer work of
Broadhurst and Kreimer~\cite{Broadhurst:1995km,Broadhurst:1996kc}.

We actually need a more general definition of abstract periods given
in~\cite{Kontsevich:2001}. Let's consider $X(\IC)$ a smooth algebraic
variety of dimension $n$ over $\mathbb Q$.  Consider 
$D\subset X$ a divisor with normal crossings, which means that locally
this is a union of coordinate hyperplanes of dimension $n-1$. Let
$\eta\in \Omega^n(X)$, and let  $\Delta\in H_n(X(\IC), D(\IC);
\IQ)$ a singular $n$-chain on $X(\IC)$ with boundary on the divisor
$D(\IC)$.
To the quadruple $(X,D,\omega,\Delta)$ we can associate a complex
number called the \emph{period of the quadruple}
\begin{equation}\label{e:defP}
  P(X,D,\omega,\Delta)= \int_\Delta \eta\,.
\end{equation}
In order that this definition is compatible with the example of
periods given previously, in particular the behaviour under change of
variables one needs to introduce the notation of equivalence classes
of  quadruples for periods leading to the same period~\eqref{e:defP}.
To this end one defines the space $\mathbf P$ of \emph{effective
  periods}  as the $\IQ$-vector space of equivalence classes modulo (a) linearity in $\eta$ and
$\Delta$, (b) under  change variables, (c) and integration by part of
Stokes formula. The map from $\mathbf P$ to the space of periods
$\mathcal P$ is clearly surjective, and it is conjectured to be
injective providing an isomorphism. We refer to the review article~\cite{Kontsevich:2001} for more details.

\section{{\bf Mixed Hodge structures for Feynman graph integrals}}
\label{sec:motive}

In this text we are focusing on ultraviolet and infrared finite
Feynman graph integrals. A discussion of 
  logarithmically divergent graphs can be found
  in~\cite{Bloch:2005bh}.

\medskip

To a Feynman graph one can associate two graph hyper-surfaces. One
defined from  the determinant of the vacuum graph period
matrix (the first Symanzik polynomial introduced in section~\ref{sec:feynm-param})  
\begin{equation}
X^0_\Gamma:=  \{ \det \Omega= \cU(x_i)=0 | x_i\in\mathbb
P^{n-1}(\IR)\}\,,
\end{equation}
 and one  graph hyper-surface $X_\Gamma$  defined by the  locus for the
zeros of the second Symanzik polynomial $\cF$ in~(\ref{e:GraphIZ2})
\begin{equation}
  X_\Gamma:=  \{  \cF(x_i)=0 | x_i\in\mathbb  P^{n-1}(\IR)\}\,.
\end{equation}
The polar part $X_\eta$ of the Feynman integral in~\eqref{e:GraphIZ2}
is the union of these two graph hyper-surfaces unless for $2\nu= (l+1)D$
when it is  only given by $X_\Gamma$ or $2\nu=lD$ when it is only given by
$X^0_\Gamma$. 
Although the integrand $\eta$ is a closed form  such that $\eta\in
H^{n-1}(\mathbb P^{n-1}\backslash X_\eta)$,  in general the domain
$\Delta$ has a boundary and therefore its  homology class is not in $H_n(\mathbb P^{n-1}\backslash
X_\eta)$.  This difficulty will be resolved by considering the relative cohomology.

In general the  polar part $X_\eta$ of the differential form 
  entering the expression of the Feynman integral
  in~\eqref{e:GraphIZ2}, intersects the boundary of the domain of
integration $\partial\Delta\cap X_\eta\neq \emptyset$. 
We need to consider a blow-up in $\IP^{n-1}$ of linear space
$f:\cP\to \IP^{n-1}$, such that all the vertices of $\Delta$ lie in
$\cP\backslash\mathcal X$ where $\mathcal X$ is the strict transform of $X_\eta$.
Let  $\mathcal B$ be the total inverse image of the coordinate simplex
$\{x_1x_2\cdots x_n=0| [x_1,\dots,x_n]\in\IP^n\}$.

As been explained by Bloch, Esnault and Kreimer in~\cite{Bloch:2005bh}
all of this lead to the mixed Hodge structure associated to the Feynman graph
\begin{equation}
  \label{e:defMotive}
  M(\Gamma):=H^{n-1}(\cP\backslash \mathcal X, \cB\backslash
  \cB\cap\mathcal X; \IQ)\,.
\end{equation}
 In the second part of this text we will give a description of the motive for particular Feynman integral. A list of
mixed Hodge structures associated with vacuum graphs can be found in~\cite{Bloch:2008,Schnetz}.
\subsection{Example: The massive one-loop triangle}
\label{sec:example:-massive-one}

As an illustration we consider the example of the mixed
Hodge structure for the massive one-loop
triangle following~\cite[section~13]{Bloch:2010gk}.

The Feynman integral is given by 
\begin{equation}
  I_\triangleright(p_1,p_2,p_3)={\mu^2\over\pi^2}\, \int {d^4\ell \over \ell^2\,
   (\ell+p_1)^2\, (\ell-p_3)^2}\,,
\end{equation}
with $p_i^2=m_i^2\neq 0$ non vanishing external masses and the
momentum conservation constraint $p_1+p_2+p_3=0$. This integral is
finite being free of ultraviolet divergences in four dimensions,  and of
infrared divergences for non vanishing  masses.

A direct application of section~\ref{sec:feynm-param}, for the case of
a graph at $l=1$ loop, with $n=3$ edges in $D=4$ dimensions, leads to the Schwinger representation 
\begin{equation}
  I_\triangleright(p_1,p_2,p_3)=\mu^2\, \int_{x_1,x_2,x_3\geq0} {
   x_1 dx_2\wedge dx_3-x_2 dx_1\wedge dx_3+x_3 dx_1\wedge dx_2\over (x_1+x_2+x_3)(x_1x_2 m^2_3+x_1x_3 m^2_2+x_2x_3m^2_1)}\,.
\end{equation}
The graph hyper-surfaces are the line $X^0_\Gamma=\{x_1+x_2+x_3=0|
x_i\in\mathbb P^2(\IR)\}$  and the conic $X_\triangleright:=\{x_1x_2 m^2_3+x_1x_3
m^2_2+x_2x_3m^2_1=0| x_i\in\mathbb P^2(\IR)\}$. The polar
part of the integrand is $X_\eta=X^0_\Gamma \cup X_\triangleright$.
The domain of integration is the triangle $\Delta=\{x_1\geq 0, x_2\geq0, x_3\geq0\}$. The intersection of the graph
polar part and the domain of integration is given by the three points
$X_\eta\cap \Delta=\{[1:0:0], [0:1:0], [0:0:1]\}$.  Let $f: \mathcal P\to \mathbb
P^2$ the blow-up of the three vertices $\{(x_1=x_2=0), (x_1=x_3=0),
(x_2=x_3=0)\}$. Let $E_i\subset \cP$ the exceptional divisors such that
$E_i$ lies over the intersection $(x_j=x_k=0)$, with $(i,j,k)$
a permutation of $(1,2,3)$. Finally, let $F_i\subset \cP$ be the strict
transform of the locus $\{x_i=0\}$.  The blown up domain of
integration is the hexagon $\mathcal B:=f^*\Delta=\cup_{i=1}^3 E_i
\cup_{i=1}^3 F_i$.  If we denote by  $\mathcal X=\mathcal L\cup
\mathcal C$ the union of the strict transform $\mathcal L$ of the line
$x_1+x_2+x_3=0$ and the strict transform $\mathcal C$ of the conic
$X_\triangleright$. The mixed Hodge
structure of the one-loop massive triangle graph in four dimension is
given by~\eqref{e:defMotive}.

\section{{\bf Variation of mixed Hodge structures}}
\label{sec:hodge-structures}

A \emph{pure Hodge structure
of weight} $n$ is an algebraic structure generalizing the
Hodge theory for compact complex manifold.
For a compact complex manifold  $\cM$ the de Rham  cohomology groups
$H^n(\cM):=H^n(\cM,\IR)\otimes \IC$ can be decomposed 
\begin{equation}
  H^n(\cM)= \bigoplus_{p+q=n} H^{p,q}(\cM), \qquad\textrm{with}\qquad  \overline{H^{p,q}(\cM)}= H^{q,p}(\cM)\,.
\end{equation}
The Dolbeaut cohomology groups  $H^{p,q}(\cM)$ are   defined
as the $\bar\partial$-closed $(p,q)$-forms modulo $\bar\partial
A^{p,q-1}(\cM)$ (see~\cite{Griffiths:1978} for a more detailed exposition).
By definition $H^n(\cM)$ is pure Hodge structure of weight $n$.

When one does not have a complex structure one defines a
pure Hodge structure  from a Hodge filtration. Let consider 
a finite dimensional $\IQ$-vector space  $H=H_\IQ$. Suppose given a decreasing filtration
$F^\bullet H_\IC$ on $H_\IC:=H_\IQ\otimes \IC$, 
\begin{equation}
  H_\IC\supseteq\cdots \supseteq F^{p-1}H_\IC \supseteq F^pH_\IC\supseteq
  F^{p+1}H_\IC\supseteq\cdots \supseteq (0)\,.  
\end{equation}
One says that $F^\bullet H_{\IC}$ defines a \emph{pure Hodge structure of weight $n$}
\begin{equation}
H_\IC^n := \bigoplus_{p+q=n}   H^{p,q}, \qquad  \textrm{where}\qquad
H^{p,q}:= F^p H_\IC\cap \overline{ F^q H_\IC}
\end{equation}
where $\overline {F^q H_\IC}$ is the complex conjugate of $F^pH_\IC$.

Pure Hodge structures are defined for smooth compact manifolds $\cM$, but 
Feynman graph integrals involve  non-compact or non-smooth
varieties which require using the generalizations provided by the
mixed Hodge structures introduced by Deligne~\cite{Deligne:1970}.

\medskip

The only pure Hodge structure of dimension one is the \emph{Tate Hodge
  structure} $\IQ(n)$ with

\begin{equation}
  F^p \IQ(n)_\IC =
  \begin{cases}
    0 &\textrm{for}~p>-n\cr
   \IQ(n)_\IC&\textrm{for}~i\leq -n\,.
  \end{cases}
\end{equation}
This means that $\IQ(n)_\IC=H^{-n,-n}(\IQ(n)_\IC)$ and $\IQ(n)$ has
weight $-2n$. Notice that $\IQ(n)\otimes\IQ(m)=\IQ(n+m)$,
  therefore $\IQ(n)=\otimes^n \IQ(1)$ for $n\in\ZZ$.

\medskip

A \emph{mixed Hodge structure} on $H$ is a pair of   (finite,
separated, exhaustive)  filtrations: (a) an increasing filtration $W_\bullet H_\IQ$
called the \emph{weight filtration}, (b) a decreasing filtration, the
Hodge filtration $F^\bullet H_\IC$  described earlier.
The Hodge structure on $H$ induces a filtration on  the graded pieces
for the weight filtration $gr_n^W H:=W_n H/W_{n-1}H$.
By definition for a mixed Hodge structure, the filtration $gr^W_n H$
should be a pure Hodge structure of weight $n$.

\medskip

A mixed Hodge structure $H$ is called \emph{mixed Tate} if
\begin{equation}
  gr^W_n H=
  \begin{cases}
    0 & \textrm{for}~n=2m-1\cr
\bigoplus \IQ(-m) &\textrm{for}~n=2m\,.
  \end{cases}
\end{equation}

\medskip

From mixed Hodge structures one can define a matrix of  periods.
For a mixed Tate Hodge structure the weight and the Hodge filtrations
are opposite since $F^{p+1}H_\IC\cap W_{2p} H_\IC=(0)$, and
$H_\IC=\bigoplus_p F^p H_\IC \cap W_{2p} H_\IC$.  We first make a
choice of a basis
$\{e^{p,p}_i\in F^pH_\IC\cap W_{2p}H_\IC\}$ of $H_\IC$.
Then expressing the basis elements $\{\varepsilon_i\}$ for $W_\bullet H$ in terms of the basis for $H_\IC$ gives a
\emph{period matrix} with columns composed by the basis elements of $F^\bullet H$.
With a proper choice of the basis for $W_\bullet H$ one can ensure that the
period matrix is block lower triangular, with the block diagonal
elements corresponding to  $gr^W_n H$ given by
$(2i\pi)^{-n}$.  In the case of the polylogarithms such a period
matrix is given in eq.~\eqref{e:MonLi}.

\medskip

Finally, we need to introduce the notation of \emph{variation of Hodge
  structure} needed to take into account that Feynman integrals lead
to families of Hodge structures parametrized by the variation of the
kinematics invariants. These concepts have been introduced by
Griffiths in~\cite{Griffiths:1968} and generalized to mixed Hodge modules over complex
varieties by M.~Saito in~\cite{Saito:1989}. We refer to these works for details
about this, but a particular case of variation of mixed Hodge
structure for  polylogarithms is discussed in the next section. 

A large class of amplitudes evaluate to (multiple) polylogarithms. In
this case a study of the  discontinuities of the amplitude can  give
access to interesting  algebraic structures~\cite{Abreu:2014cla}.
As well elliptic integrals arise from multiloop amplitudes~\cite{Caffo:1998du,MullerStach:2011ru,CaronHuot:2012ab,Adams:2013nia,Bloch:2013tra,Remiddi:2013joa} .
One example is the sunset Feynman integral studied in
section~\ref{sec:sunsetintegral}. The value of the integral  is obtained from a variation of mixed
Hodge structure when the external momentum is varying~\cite{Bloch:2013tra}.

\subsection{Polylogarithms}
\label{sec:polylogs}

A  very clear motivic approach to  polylogarithms is detailed in the
article by Beilinson and Deligne~\cite{DeligneBeilinson}. We only
refer to the main points needed for the present discussion, for
details we refer to the articles~\cite{Hain,DeligneBeilinson}.
The iterated integral definition of the polylogarithms 
\begin{eqnarray}
  \Li1(z)&:=&-\log(1-z)=\int_0^z {dt\over 1-t}\cr
\Li{k+1}(z)&:=& \int_0^z \,\Li{k}(z) \,{dt\over t},\qquad k\geq1
\end{eqnarray}
imply that they provide multivalued function on $\mathbb
P^1\backslash\{0,1,\infty\}$.  These multivalued function have
monodromy properties. To this end defined the lower triangular matrix
of size $n\times n$ as
\begin{equation}\label{e:MonLi}
    A(z):=
    \begin{pmatrix}
      1&0&\cdots &&&\cdots &0\cr
   -\Li1(z)& 1&0&\cdots&&\cdots & 0\cr
-\Li2(z)&  \log z& 1&0&\cdots&\cdots&0\cr
-\Li3(z)&  {(\log z)^2\over 2!}& \log z&1&0&\cdots &0\cr
\vdots&     \vdots& &\ddots &\ddots&&\vdots
    \end{pmatrix}\, \textrm{diag}(1,2i\pi,\dots,(2i\pi)^n)\,.
\end{equation}
so that $A_{1k}(z)= -\Li{k}(z)$ for $1\leq k\leq n$, and $A_{pq}(z)=
(2i\pi)^{p-1} (\log z)^{q-p}/(q-p)!$ for $2\leq p<q\leq n$.

For a fixed value of $z$ this matrix is the period matrix associated
with the mixed Hodge structure for the polylogarithms, the columns are
the weight and the lines are the Hodge degree. 

A determination of this matrix $A(z)$ depends on the path $\gamma$ in
$\mathbb P^1\backslash\{0,1,\infty\}$ and a point $z\in]0,1[$. For a
counterclockwise path  $\gamma_0$ around 0 or $\gamma_1$
around 1 the determination of $A(z)$ is changed as
\begin{equation}
  A_{\gamma\gamma_i}(z)= A_\gamma(z) \,\exp(e_i);\qquad i=0,1
\end{equation}
where $e_i$ are the nilpotent matrices  
\begin{equation}
  e_0:=
  \begin{pmatrix}
    0 &0&\cdots &\cdots&0\cr
    0 &1&\cdots &\cdots&0\cr
    0 &0&1&\cdots &0\cr
    0&\cdots&0&\ddots&0
  \end{pmatrix};
\quad
  e_1:=
  \begin{pmatrix}
    0 &0&\cdots &0\cr
    1 &0&\cdots &0\cr
    0 &\cdots&\ddots&\cdots
  \end{pmatrix}\,.
\end{equation}
The matrix $A(z)$ satisfies the differential equation
\begin{equation}\label{e:KZ}
  d A(z)= \left(e_0 \, d\log(z)+ e_1 d\log(z-1)  \right)\, A(z)\,.
\end{equation}
This  differential equation defined over $\IC\backslash\{0,1\}$
defined the \emph{$nth$ polylogarithm local system}. This local system 
underlies a good variation of mixed Hodge structure whose weight graded quotients are canonically isomorphic to
$\IQ,\IQ(1),\dots,\IQ(n)$~\cite[theorem~7.1]{Hain}.

\medskip
One can define   single-valued real analytic function on $\mathbb P^1(\mathbb
C)\backslash\{0,1,\infty\}$, and continuous on $\mathbb P^1(\mathbb
C)$. 
The first important example is the  \emph{Bloch-Wigner dilogarithm}
defined as~\cite{BlochCMR,ZagierHouches} 
\begin{equation}\label{e:Ddef}
  D(z):=\Imm\left(\Li2(z)+\log|z| \log(1-z)\right).  
\end{equation}
The Bloch-Wigner dilogarithm function satisfies the following
functional equations
\begin{eqnarray}
  \label{e:Dfunc}
  D(z)&=&-D(\bar z)=D(1-z^{-1})=D((1-z)^{-1})\cr
&=&-D(z^{-1})=-D(1-z)=-D(-z(1-z)^{-1})\,.
\end{eqnarray}
The differential of the Bloch-Wigner dilogarithm $D(z)$ is given by 
\begin{equation}
\label{e:dD}
  dD(z)= \log|z| d\arg (1-z) - \log|1-z| d\arg(z)\,.
\end{equation}
At higher-order there is no unique form for the real analytic version
of the polylogarithm. 
A particularly nice version with respect to Hodge structure
 provided by Beilinson and Deligne
in~\cite{DeligneBeilinson} is given by
\begin{equation}\label{e:Lmdef}
  \mathcal L_m(z):=\sum_{k=0}^{m-1} {B_k\over k!} \, (\log(z\bar z))^k \times
\begin{cases}
  \Ree(\Li{m-k}(z))& \textrm{for}~m =1\mod 2\cr
\Imm(\Li{m-k}(z))& \textrm{for}~m=0\mod 2\,.
\end{cases}
\end{equation}
where $B_k$ are Bernoulli numbers $x/(e^x-1)=\sum_{k\geq0} B_k \,x^k/k!$.

\medskip
A \emph{dilogarithm Hodge structure}, relevant to one-loop  amplitudes
in four dimensions, has been defined
in~\cite{Bloch:2010gk} as a mixed Tate Hodge structure such that for
some integer $n$,  $gr^W_{2p}H=(0)$ for $p\neq n,n+1,n+2$.
\subsection{Polylogarithms and Feynman integrals}
\label{sec:polyl-feynm-integr}

The construction of the mixed Hodge structure of the massive triangle
in section~\ref{sec:example:-massive-one} was shown
in~\cite{Bloch:2005bh,Bloch:2010gk} to correspond to the dilogarithm
Hodge structure describe above.

\medskip

Are all the Feynman integral expressible as polylogarithms or multiple
polylogarithms in several variables?
It is conjectured in~\cite{Henn:2013pwa} that using integration by
parts one could always express Feynman integrals as a combination of a
finite set of master integrals satisfying the differential
equation~\eqref{e:KZ}, therefore leading to multiple polylogarithm functions.
Various high-loop graphs have been shown to evaluate to multiple polylogarithms~\cite{Henn:2013tua,Henn:2013woa,Caron-Huot:2014lda,Abreu:2014cla,Panzer:2014gra}.

\medskip

Counter-examples leading to elliptic functions are known in the massless
case~\cite{Brown:2012ia,CaronHuot:2012ab,Nandan:2013ip} or the massive
case by 
the sunset graph~\cite{Laporta:2004rb,MullerStach:2011ru,Adams:2013nia,Adams:2014vja}
and three-loop banana graph with all equal internal masses~\cite{BlochKerrVanhove}.  
The banana graphs with all equal internal masses, discussed in
section~\ref{sec:banana}, lead to elliptic polylogarithms discussed below.

\medskip 
Different functions are expected from other classes of
Feynman graphs. Determining and evaluating Feynman integrals are
open and difficult questions, where one can hope  that mixed Hodge structures or
motivic methods be useful. 

\section{{\bf Elliptic polylogarithms}}\label{sec:ellipticdilog}

In this section we recall the main properties of the elliptic
polylogarithms following~\cite{BlochCMR,ZagierElliptic,GanglZagier,Levin,BL1}.

Let $\mathcal E(\IC)$ be an elliptic curve over $\IC$.
The elliptic curve can be viewed either as the complex plane modded by a
two-dimensional lattice $\mathcal E(\IC)\cong\IC/(\ZZ
\varpi_1+\ZZ\varpi_2)$. A point $z\in  \IC/(\ZZ
\varpi_1+\ZZ\varpi_2)$ is associated to
a point $P:=(\wp(z),\wp'(z))$ on $\mathcal E(\IC)$
where $\wp(z)=z^{-2}+\sum_{(m,n)\neq(0,0)} \left((z+m\varpi_1+n\varpi_2)^{-2}- (m\varpi_1+n\varpi_2)^{-2}\right)$ is the Weierstra\ss{} function and  $q:=\exp(2i\pi \tau)$ with 
$\tau=\varpi_2/\varpi_1$ the period ratio in the upper-half plane $\mathbb H=\{\tau| \Ree(\tau)\in
\IR, \Imm(\tau)>0\}$.
Or  we can  see the elliptic curve as   $\mathcal
E(\IC)\cong\IC^\times/q^\ZZ$. A point $P$ on the elliptic curve is then mapped to $x:=e^{2i\pi z}$.

\medskip

One defines an \emph{elliptic polylogarithm}
$\mathcal L_n^{\mathcal E} : \mathcal E(\IC)\to \IR$ as the average
of the real unvalued version of the polylogarithms
\begin{equation}
  \label{e:DlogDef}
  \mathcal L_m^{\mathcal E}(P):= \sum_{n\in\ZZ}  \mathcal L_m(x \, q^n) 
\end{equation}
where $q:=\exp(2i\pi\tau)$ with $\tau\in \mathfrak h:=\{\tau | \Ree(\tau)\in\IR, \Imm(\tau)>0\}$.
This series converges absolutely with exponential decay and is
invariant under the transformation $x\mapsto qx$ and $x\mapsto q^{-1}x$.

\medskip

If we have a collection of points $P_r$ on the elliptic curve one can
consider a  linear combination of the elliptic polylogarithms.
Such objects  play an important role when computing regulators for
elliptic curves, and in the so-call Beilinson conjecture
relating the value of the regulator map to the value of  $L$-function
of the elliptic curve~\cite{Beilinson,Denninger,BlochCMR,Soule,Brunault}.

\medskip
Interestingly, as explained in~\cite{Bloch:2013tra}, elliptic
polylogarithms from Feynman graphs differ from~\eqref{e:DlogDef}.
A simple physical reason is that the Feynman integral is a
multivalued function therefore cannot be build from a real analytic
version of the polylogarithms.  
For the examples discussed in section~\ref{sec:banana} we will need
the following sums of the elliptic dilogarithms 
\begin{equation}
\sum_{r=1}^{n_r} c_r \sum_{n\geq0} \Li2(q^n z_r)
\end{equation}
where $z_r$ is a finite set of points on the elliptic curve and $c_r$
are rational numbers.  This expression is invariant under $z\mapsto qz$
and $z\mapsto q^{-1}z$ only for a \emph{very
special} choice of set of points depending on the (algebraic) geometry of the graph.
A more precise definition of the quantity appearing from the two-loop
sunset Feynman graph is given in equation~\eqref{e:Esunset}.

\subsection{Mahler measure}
\label{sec:mahler-measure}

A \emph{logarithmic Mahler measure} is defined by 
\begin{equation}
  \mu(F) := \oint_{|x_1|=\cdots=|x_n|=1} \log| F(x_1,\dots,x_n)| \,\prod_{i=1}^n
  {dx_i\over   2i\pi x_i}\,,
\end{equation}
and the \emph{Mahler measure} is defined by $M(F):=\exp(\mu(F))$.
In the definition $F(x_1,\dots,x_n)$ is a Laurent polynomial in
$x_i$.

Numerical experimentations by Boyd~\cite{Boyd} pointed out to a
relation between the logarithmic Mahler measure for certain Laurent
polynomials $F$ and values of $L$-functions of the projective plane
curve $C_F:  F(x_1,\dots,x_n)=0$
\begin{equation}\label{e:MtoL}
  \mu(F) = \IQ^\times \, L'(Z_F,0)\,.  
\end{equation}

  In~\cite{RV1} (see as well~\cite{BoydI,BoydII}) Rodrigez-Villegas
  showed that the logarithmic Mahler  measure is given by evaluating
  the Bloch regulator leading to expressions given by the
  Bloch-Wigner dilogarithm. The   relation in~\eqref{e:MtoL} is then a consequence of 
the conjectures by Bloch~\cite{BlochCMR} and
Beilinson~\cite{Beilinson} relating regulators for elliptic curves to
the values of $L$-functions (see~\cite{Soule,Brunault} for some review
on these conjectures). 

\medskip

Let consider the {logarithmic Mahler measure}  defined using the
second Symanzik polynomial $\cF_2(x,y;t)=(1+x+y)(x+y+xy)-txy$ for the
two-loop $n=3$ banana graph of figure~\ref{fig:banana}
\begin{equation}
    \mu_\circleddash(t)= {1\over (2i\pi)^2}\int_{|x|=|y|=1}  \log(|\cF_2(x,y;t)|)
    {dxdy\over xy}\,.
\end{equation}
The Mahler measure associated with this
polynomial has been studied by
Stienstra~\cite{Stienstra:2005wy,Stienstra:2005wz} and Lalin-Rogers in~\cite{RogerL}.

Consider the field $F=\IQ(\cEs)$ where $\cEs=\{(x,y)\in\IP^2 |
\cF_2(x,y;t)=0\}$ is the sunset elliptic curve, and consider a N\'eron $\widehat{\cEs}$
model of $\cEs$ over $\ZZ$.   The regulator map is
an application from the higher regulator $K_2(\widehat{\cEs})$ to $H^1(\cEs,\IR)$~\cite{BlochCMR}.
The regulator map is defined by
\begin{eqnarray}\label{e:DefReg}
    r:  K_2(\cEs) &\to&  H^1(\cEs,\IR)\cr
    \{x,y\}&\mapsto&  \{\gamma \to \int_\gamma \eta(x,y)\}\,,
\end{eqnarray}
where
\begin{equation}
  \eta(x,y)= \log|x| d\arg (y)-\log|y| d\arg x  \,.
\end{equation}
Notice that $  \eta(x,1-x)=dD(x)  $ the differential of the
Bloch-Wigner dilogarithm.

If $x$ and $y$ are non-constant function on $\cEs$ with divisors
$(x)=\sum_i x_i (a_i)$ and $(y)=\sum_i n_i (b_i)$ one associates the
quantity $(x)\diamond (y)=\sum_{i,j} m_i n_i (a_i-b_j)$ 

A theorem by Beilinson states that if $\omega\in \Omega^1(\cEs)$ then 
\begin{equation}
  \int_{\cEs(\IC)} \omega\wedge \eta(x,y)=  \varpi_r\,  R_\tau(
  (x)\diamond (y))  
\end{equation}
where $R_\tau(z)$ is the Kronecker-Eisenstein series~\cite{weil,BlochCMR} defined as
\begin{equation}
  \label{e:Rtau}
  R_\tau(e^{2i\pi(a+b\tau)}):=
  {\Imm(\tau)^2\over\pi^2}\,\sum_{(p,q)\neq(0,0)} {e^{2i\pi (aq-pb)}\over (p+q\tau)^2(p+q\bar\tau)}\,.
\end{equation}

The logarithmic Mahler measure for the sunset graph  is expressed as a sum of
elliptic-dilogarithm evaluated at torsion points on the elliptic
curve
\begin{equation}
  \mu_\circleddash(t)= -3\,\Imm \left(R_\tau(\zeta_6)+R_\tau(\zeta_6^2)\right)
\end{equation}
with $\zeta_6=\exp(i\pi/3)$ is a sixth root of unity
and $\tau=\varpi_2/\varpi_1$ is the period ratio of the elliptic curve.
This relation is true for $t$ large enough so that the elliptic curve $\cEs$
does not intersect the torus $\mathbb T^2=\{|x|=|y|=1\}$.

The Beilinson conjecture~\cite{Beilinson,Soule,Brunault} implies that the Mahler measure is
rationally related to the value of the Hasse-Weil $L$-function for the
sunset elliptic curve evaluated at $s=2$
\begin{equation}
  \label{eq:7}
  \mu_\circleddash(t) = \IQ^\times\, L(\cEs(t),2)\,.
\end{equation}
which can be easily numerically checked using~\cite{sage}.

Differentiating the Mahler measure with respect to $t$ gives

\begin{equation}\label{e:periodsunset}
 g_1(t):=-t {d\mu_\circleddash(t)\over dt}=
 {1\over(2i\pi)^2}\oint_{|x|=1}\oint_{|y|=1}  {t\, dxdy\over \cF_2(x,y;t)}\,. 
\end{equation}
This quantity is actually a period of the elliptic
curve~\cite{Stienstra:2005wy} since we are integrating the two-form
$\omega={t\, dxdy\over \cF_2(x,y;t)}$ over a two-cycle given by the
torus $\mathbb T^2=\{|x|=1, |y|=1\}$ (for $t$ large enough so that the
elliptic curve does not intersect the torus).

\specialsection*{{\scshape The banana integrals
  in two dimensions}}

In this section we will discuss the two-point $n-1$-loop  all equal mass banana
graphs in two dimensions.

We first provide an algorithm for determining  the differential equation satisfied by these
amplitudes to all loop order, and present the solution for the
one-loop banana (bubble) graph and the two-loop banana (sunset)
graph given in~\cite{Bloch:2013tra}.  The solution to the three-loop
banana graph will appear in~\cite{BlochKerrVanhove}.

\section{{\bf Schwinger representation}\label{sec:banana}}
\label{sec:schw-repr}

We look at the $n-1$-loop banana graph of figure~\ref{fig:banana}
evaluated in $D=2$ euclidean dimensions 
\begin{equation}
  I_n^2(m_1,\dots,m_n;K)= \int_{\IR^{2n}} {\prod_{i=1}^nd^2\ell_i  \delta^{(2)}(\sum_{i=1}^n\ell_i=K)\over
    \prod_{i=1}^n (\ell_i^2+m_i^2)}\,,
\end{equation}
where $\ell_i$ for $1\leq i\leq n$ are the momenta of each propagator.
The steps described in section~\ref{sec:feynm-param} lead to the
following representation of the banana integrals in two dimensions
\begin{equation}
  I_n^2= \int_{x_i\geq0} {\delta(x_n=1) \over \cF_n}\,\prod_{i=1}^ndx_i
\end{equation}
with
\begin{equation}
  \cF_n= (\sum_{i=1}^n x_i m_i^2) \cU_n - K^2\prod_{i=1}^n x_i
\end{equation}
where  $\cU_n$ is the determinant of the period matrix $\Omega$ given in~\eqref{e:OmegaBanana}
\begin{equation}
  \cU_n=\prod_{i=1}^n x_i \, \left(\sum_{i=1}^n {1\over x_i}\right)\,.
\end{equation}
In order to determine the differential equation satisfied by the all
equal mass banana graphs we provide an alternative expression for the
banana integrals. 
For  $K^2<(\sum_{i=1}^n m_i)^2$,   one can perform a series 
expansion 
\begin{eqnarray}
  I_n^2&=& \int_{[0,+\infty[^{n-1}} \, {\delta(x_n=1)\over    \sum_{i=1}^n m_i^2
    x_i\sum_{i=1}^n x_i^{-1} -t  }  \, \prod_{i=1}^{n-1}
  {dx_i\over x_i}\cr
&=&\sum_{k\geq0}  t^k  \int_{[0,+\infty[^{n-1}} \, {\delta(x_n=1)\over    (\sum_{i=1}^n m_i^2
    x_i)^{k+1}(\sum_{i=1}^n x_i^{-1})^{k+1}  } \, \prod_{i=1}^{n}
  {dx_i\over x_i}\,.
\end{eqnarray}
Exponentiating the denominators 
\begin{equation}
  I_n^2=\sum_{k\geq0} { t^k\over k!^2}  \int_{[0,+\infty[^{n+1}}\delta(x_n=1)
  \,e^{-u   (\sum_{i=1}^n m_i^2
    x_i)- v (\sum_{i=1}^n x_i^{-1})} \, {dudv\over u^{-k} v^{-k}}\prod_{i=1}^{n}
  {dx_i\over x_i}\,.
\end{equation}
Using the integral representation for the  $K_0$ Bessel function 
\begin{equation}
  \int_0^{+\infty}  e^{-u m^2 x - {v\over x}} \, {dx\over x}= 2
  K_0(2m \sqrt{uv})
\end{equation}
one gets
\begin{equation}
  I_n^2=2^{n-1}\sum_{k\geq0} { t^k\over k!^2} \, \prod_{i=1}^{n-1} K_0(2m_i\sqrt{uv}) \int_{[0,+\infty[^{2}}
  \,e^{-u  m_n^2- v} \, u^k v^k  dudv\,.
\end{equation}
Now setting $uv=(x/2)^2$, the integral over $v$ gives a $K_0$ Bessel
function with the result
\begin{equation}
  I_n^2=2^{n-1}\int_{[0,+\infty[}\sum_{k\geq0} \left(\sqrt t x\over
      2\right)^{2k}{1\over k!^2} \, \prod_{i=1}^{n} K_0(m_ix)  dx\,.
\end{equation}
Using the series expansion of the $I_0$ Bessel function 
\begin{equation}
  I_0(x)= \sum_{k\geq0} \left(x\over2\right)^{2k}\,{1\over k!^2}
\end{equation}
we get the following representation for the banana graph (see~\cite{Bailey:2008ib}
for a previous appearance of this formula at the special values
$K^2=m^2$ and all equal masses $m_i=m$)
\begin{equation}
  \label{e:BanananBessel}
  I_n^2=2^{n-1}\int_0^{+\infty} x \, I_0(\sqrt{K^2}x) \,\prod_{i=1}^n
  K_0(m_ix)\, dx\,.
\end{equation}
%

\section{{\bf The differential equation for the banana graphs at all
    loop orders}}
\label{sec:all-n-differential}

We derive a differential equation for the $n-1$-loop all equal mass
banana graphs
\begin{equation}
  I_n^2(t):= 2^{n-1}\int_0^\infty x \,I_0(\sqrt t x)\, K_0(x)^n dx\,.  
\end{equation}
This will generalize to all loop order the differential equations given for
the two loops case in~\cite{Laporta:2004rb,MullerStach:2011ru,Adams:2013nia}
We first prove the existence of a differential equation for the
integral. In~\cite{BorweinSalvy}  it is proven that the Bessel
function $K_0(x)$ satisfies the differential equation
\begin{equation}\label{eq:diffK0}
  L_{n+1} K_0(x)^n=0  
\end{equation}
where $L_{n+1}$ is a degree $n+1$ differential operator expressed as a
polynomial in the differential operator $\theta_x= x{d\over
  dx}$  of the form
$L_{n+1}=\theta_x^{n+1} + \sum_{k=0}^n p_k(x^2) \theta_x^k $.
This operator is obtained by the recursion given
in~\cite{BorweinSalvy}
\begin{eqnarray}\label{e:recuLn}
    L_1&=&\theta_x\\
   L_{k+1}&=&\theta_x L_k-x^2 k (n+1-k) L_{k-1},\qquad 1\leq k\leq n\,.
\end{eqnarray}
Setting $\theta_t:=t{d\over   dt}$ we have the following relations
\begin{eqnarray}
2  \theta_t I_0(\sqrt t x)&=& \theta_x I_0(\sqrt t x);\\ 
 \theta_t^2 I_0(\sqrt t x)&=& t \left(x\over 2\right)^2 \, I_0(\sqrt t x);\\
(\theta_t^3-\theta_t^2) I_0(\sqrt t x)&=&{tx^2\over 8}
\theta_x(I_0(\sqrt t x))\,,
\end{eqnarray}
and integrating by parts, one can
convert the differential~\eqref{eq:diffK0} into a differential
equation for $\hat L_{n+1} I_n^2(t)=0$ for the banana integral $I_n^2(t)$.  
Since
\begin{equation}
  \int \, x^k f(x) \theta_x g(x)\,dx= -(k+1)\int x^k f(x)g(x) -\int
  x^k f(x) g(x)\,dx  \,.
\end{equation}
The polynomial coefficients  $p_k(x)$ in~\eqref{eq:diffK0} are
polynomials in $x^2$ such that
$p_k(0)=0$~\cite{BorweinSalvy}. Therefore the differential operator $\hat
L_{n+1}= \theta_t^2\, \tilde L_{n-1}$ where $\tilde L_{n-1}$ is a
differential operator of at most degree $n-1$. 
We conclude that the integral satisfies the differential equation 
\begin{equation}
\tilde L_{n-1} I_n^2(t)= S_n + \tilde S_n\,\log(t)\,.  
\end{equation}
where $S_n$ and $\tilde S_n$ are constants.  It is easy to check that
$I_n^2(t)$ has a finite value of $t=0$. Therefore $\tilde S_n=0$ and
the $n-1$-loop banana integral satisfy a  differential equation of
order $n-1$ with a constant inhomogeneous term.

The differential operator acting on  the integral $I_n^2(t)$ is given by
\begin{equation}\label{e:defLn}
\tilde L_{n-1}=  \sum_{k=0}^{n-1} q_k(t) \, {d^{k}\over dt^{k}}
\end{equation}
with the coefficient $q_k(t)$ polynomials of degree  $k+1$ for $0\leq
k\leq n-1$.
The top degree and lowest-degree polynomial are given by 
\begin{eqnarray}\label{e:qnCoeff}
  q_{n-1}(t)&=&t^{\left\lfloor
      n\over2\right\rfloor+\eta(n)}\,\prod_{i=0}^{\left\lfloor
      n\over2\right\rfloor} (t-(n-2i)^2)\\  
q_{n-2}(t)&=&{n-1\over 2}\,{dq_{n-1}(t)\over dt}\\
q_0(t)&=&t-n\,.
\end{eqnarray}
with $\eta(n)=0$ if $n\equiv 1 \mod 2$ and 1 if $n\equiv 0 \mod2$.
The inhomogeneous term $S_n$ is a constant given by 
\begin{equation}
  S_n=\int_0^\infty 2^{n-1} \, x\, \left(\sum_{k=0}^{\left\lfloor n\over
      2\right\rfloor-\eta(n)} q_k(0) \left(x\over2\right)^{2k}\right)\, K_0(x)^n dx   
\end{equation}
Numerical  evaluations for  various loop orders  give that $S_n=-n!$.

\medskip

In section~\ref{sec:maplecode} we provide  {\tt Maple} codes for generating the
differential equations  for the all equal mass banana integrals
\begin{equation}\label{e:BananaDiff}
  \left( \sum_{k=0}^{n-1} q_k(t) \, {d^{k}\over dt^{k}}\right)\,I_n^2(t)= -n!\,.
\end{equation}

\begin{table}[h]
\begin{tabular}{||c|c||}
\noalign{\hrule}
\# loops$=n-1$ & differential equation\\
\noalign{\hrule}
    $n=2$&$\left( t-2 \right) f \left( t \right) +
    t(t-4)\,f^{(1)}(t)=- 2!$\\[1ex]
$n=3$&$\left( t-3 \right) f \left( t \right) + \left(
  3\,{t}^{2}-20\,t+9 \right)f^{(1)}(t) + t (t - 1) (t -
9)f^{(2)}(t)=-3!$\\[1ex]
$n=4$&$\left( t-4 \right) f \left( t \right) + \left( 7\,{t}^{2}-68\,t+64
\right)f^{(1)}(t) + \left( 6\,{t}^{3}-90\,{t}^{2}+192\,t
\right)f^{(2)}(t)$\cr
&$+ t^2  (t - 4) (t - 16) f^{(3)}(t) =-4!$\\[1ex]
$n=5$&$\left( t-5 \right) f \left( t \right) +  (3 t - 5) (5 t -
57)f^{(1)}(t) + \left( 25\,{t}^{3}-518\,{t}^{2}+1839\,t-450
\right)f^{(2)}(t)$\cr
&$+ \left( 10\,{t}^{4}-280\,{t}^{3}+1554\,{t}^{2}-900\,t
\right)f^{(3)}(t) + t^2  (t - 25) (t - 1) (t - 9)f^{(4)}(t) =-5!$\\[1ex]
$n=6$&$\left( t-6 \right) f \left( t \right) + \left(
  31\,{t}^{2}-516\,t+1020 \right)f^{(1)}(t) + \left(
  90\,{t}^{3}-2436\,{t}^{2}+12468\,t-6912 \right) f^{(2)}(t)$\cr 
&$+ \left( 65\,{t}^{4}-2408\,{t}^{3}+19836\,{t}^{2}-27648\,t
\right)f^{(3)}(t) $ \cr
&$+ \left(
  15\,{t}^{5}-700\,{t}^{4}+7840\,{t}^{3}-17280\,{t}^{2}
\right)f^{(4)}(t)+  t^3  (t - 36) (t - 4) (t - 16) f^{(5)}(t)=-6!$\cr
\noalign{\hrule}
  \end{tabular}
  \caption{Examples of the differential equations satisfied by the all equal mass
    $n-1$-loop banana graph up to $n=6$ generated  with the {\tt Maple} code
  of section~\ref{sec:maplecode}. We made use of the notation $f^{(n)}(t):={    d^nf(t)\over dt^n}$.}
  \label{tab:bananadiff}
\end{table}

\subsection{Maple codes for the differential equations}
\label{sec:maplecode}

A derivation of the differential equations for the banana graph can be obtained  using {\sl
  Maple} and the routines  {\tt compute\_Q}
and {\tt rec\_Q} from the paper~\cite{BorweinSalvy} together with the package {\tt gfun}~\cite{gfun}.

For $t<n^2$ the integral converges and we can 
perform the series expansion
\begin{equation}
  I_n^2(t) =\sum_{k\geq0} t^k \, J_n^k  \,,
\end{equation}
where we have introduced the Bessel moments
\begin{equation}\label{e:Jnk}
  J_n^k = {2^n\over \Gamma(k+1)^2}\,
  \int_0^{+\infty}\,\left(x\over2\right)^{2k+1} \, K_0(x)^n\, dx\,.
\end{equation}

Using the result of~\cite{BorweinSalvy} on the recursion relations
satisfied by these Bessel moment $c_{n,2k+1}=2^{2k+1-n}\Gamma(k+1)^2\, J_n^k$ we deduce that these moments
satisfy a recursion relation for $k\geq0$
\begin{equation}\label{e:JnkRec}
  (k+1)^{n-1} \, J_n^k + \sum_{1\leq \ell \leq \left\lfloor
    n\over2\right\rfloor} \, P_{n,2\ell}(k) \, J_n^{k+\ell}=0
\end{equation}

A first step is to construct the recursion relations satisfied by the
coefficients $J_n^k$. For this we use the routines from~\cite{BorweinSalvy}
\begin{verbatim}
compute_Q:=proc(n,theta,t)
local k, L;
    L[0]:=1; L[1]:=theta;
    for k to n do
        L[k+1]:=expand(series(
            t*diff(L[k],t)+L[k]*theta-k*(n-k+1)*t^2*L[k-1],
            theta,infinity))
    od;
    series(convert(L[n+1],polynom),t,infinity)
end:

rec_c:=proc(c::name,n::posint,k::name)
local Q,theta,t,j;
    Q:=compute_Q(n,theta,t);
    add(factor(subs(theta=-1-k-j,coeff(Q,t,j)))*c(n,k+j),j=0..n+1)=0
end:
\end{verbatim}

The recursion relation~\eqref{e:JnkRec}, for the $J_n^k$ in eq.~\eqref{e:Jnk}, is then obtained by the routine
\begin{verbatim}
Brec:=proc(n::posint)  local e1,e2,Jnktmp,vtmp,itmp;
    Jnktmp:= (n, k) -> 2^(-n+2*k+1)*factorial(k)^2*J(k):
    e1 := subs([k = 2*K+1], rec_c(c, n, k)):
    e2 := subs([c(n, 2*K+1) = Jnktmp(n, K)], e1):
    for itmp from 1 to ceil(n/2) do 
    e2:=subs([c(n,2*K+1+2*itmp)=Inktmp(n,K+itmp)],e2) od:
    vtmp:=seq(j(K+i),i=0..ceil(n/2)):
    collect(simplify((-1)^(n-1)*e2/(4^(K+1)*factorial(K+1)^2)),{vtmp},factor)
    end:

\end{verbatim}

Then the differential equation is obtained  from the previous
recursion relation using the command {\tt rectodiffeq} from the
package~{\tt gfun}~\cite{gfun}
\begin{verbatim}
with(gfun):
rectodiffeq({Brec(n),seq(J(k)=j(n,k),k=0..floor(n/2)-1+(n mod 2))},J(K),f(t));
\end{verbatim}


\subsection{The Picard-Fuchs equation of Feynman graphs}
\label{sec:some-formal-solution}

Feynman graph hyper-surfaces lead to
Calabi-Yau geometries. It is therefore not  surprising that
 the Picard-Fuchs operators acting on the Feynman integrals are
similar to the one arising in the context of open mirror
symmetry discussed  in~\cite{MorrisonWalcher,JeffWal} for instance.  
The inhomogeneous term is however different because Feynman integrals
lead to different extensions of mixed Hodge structure than the one
encountered in the mirror symmetry case.  

\medskip

For the banana graphs we give a formal solution to the Picard-Fuchs equations
given in the previous section. Let
$\{y_1(t),\dots, y_{n-1}(t)\}$ solutions of the homogeneous
Picard-Fuchs equation $\tilde L_n(t) y_i(t)=0$ with $1\leq i\leq
n-1$ of equation~\eqref{e:defLn}, and the generalized Wronskian
\begin{equation}
  W(x,t) =
\det  \begin{pmatrix}
    y_1(x)& y_2(x) &\cdots  & y_{n-1}(x)\cr
    y'_1(x)& y'_2(x) &\cdots  & y'_{n-1}(x)\cr
\vdots & \cdots &\vdots \cr
 y^{(n-3)}_1(x)& y^{(n-3)}_2(x) &\cdots  & y^{(n-3)}_{n-1}(x)\cr
    y_1(t)& y_2(t) &\cdots  & y_{n-1}(t)\cr
  \end{pmatrix}
\end{equation}
Clearly  $\partial_t^i \left. W(x,t)\right|_{x=t}=0$ for $0\leq i\leq
n-2$, and   $\partial^{n-1}_t\left. W(x,t)\right|_{x=t}=\exp(-\int^t q_{n-2}(x)/q_{n-1}(x) dx)=
W_0/q_{n-1}^{n-1\over2}(t)$ is the Wronskian of 
the differential Picard-Fuchs operator $\tilde L_n(t)$
in~\eqref{e:defLn}. With a convenient choice of homogeneous
solutions one
can set the constant of integration  $W_0=1$.
A  formal solution to the banana Picard-Fuchs reads
\begin{equation}
  I^2_n(t) = \sum_{i=1}^{n-1} \alpha_i \, y_i(t)-n!\int_0^t
  W(x,t)\, q_{n-1}^{n-3\over2}(x)\, dx\,,
\end{equation}
where $\alpha_i$ are constant of integrations. This method has been
used~\cite{Bloch:2013tra,BlochKerrVanhove} to solve the differential equation for $n=2$ and $n=3$. 
In the following we  describe the solution of the lowest order
banana graphs at one- and two-loop orders. For a  detailed discussion of
the three-loop banana graphs we refer to~\cite{BlochKerrVanhove}.

\section{{\bf Some explicit solutions for the all equal masses banana graphs}}\label{sec:explicit-banana}
In~\cite{BroadhurstProc}  Broadhurst  provided a mixture of proofs and
numerical evidences that up to and including four loops the
special values $t=K^2/m^2=1$ for the all equal mass banana graphs
are given by values of $L$-functions.  

For generic values of  $t=K^2/m^2\in[0,(n+1)^2]$, the solution is expressible as an
elliptic dilogarithm at $n=2$ loops order~\cite{Bloch:2013tra} and
elliptic trilogarithm at $n=3$ loops
order~\cite{BlochKerrVanhove}. The situation at higher-order is not
completely clear.

In the following we present the one- and two-loop order solutions.
\subsection{The massive one-loop bubble}
\label{sec:massive-bubble}

In $D=2$ dimensions the one-loop banana graph, is the massive bubble,
which evaluates to
\begin{equation}
  I^2_2(m_1,m_2,K^2)= {\log(z^+)- \log(z^-)\over \sqrt{\Delta}}
\end{equation}
where 
\begin{eqnarray}
  z^\pm &=& (K^2-m_1^2-m_2^2\pm \sqrt{\Delta})/(2m_1^2)\cr  
\Delta_\circ&=& (K^2)^2+m_1^4+m_2^4- 2(K^2 m_1^2+K^2 m_2^2+m_1^2m_2^2)
\end{eqnarray}
where $\Delta_\circ$ is the discriminant of the equation
\begin{equation}
    (m_1^2x+m_2^2)(1+x)-K^2x=m_1^2(x-z^+)(x-z^-)=0
\end{equation}
In the single mass case $m_1=m_2=m_3=1$ the integral reads
\begin{equation}
  I_2^2(t)=- 4\, {\log\left(\sqrt t+\sqrt{t-4}\right)+\log 2\over \sqrt{t(t-4)}}  \,.
\end{equation}
This expression satisfies the differential equation for $n=2$ in table~\ref{tab:bananadiff}.
\subsection{The sunset integral}
\label{sec:sunsetintegral}

\begin{figure}[ht]
  \centering
  \includegraphics[height=5.5cm]{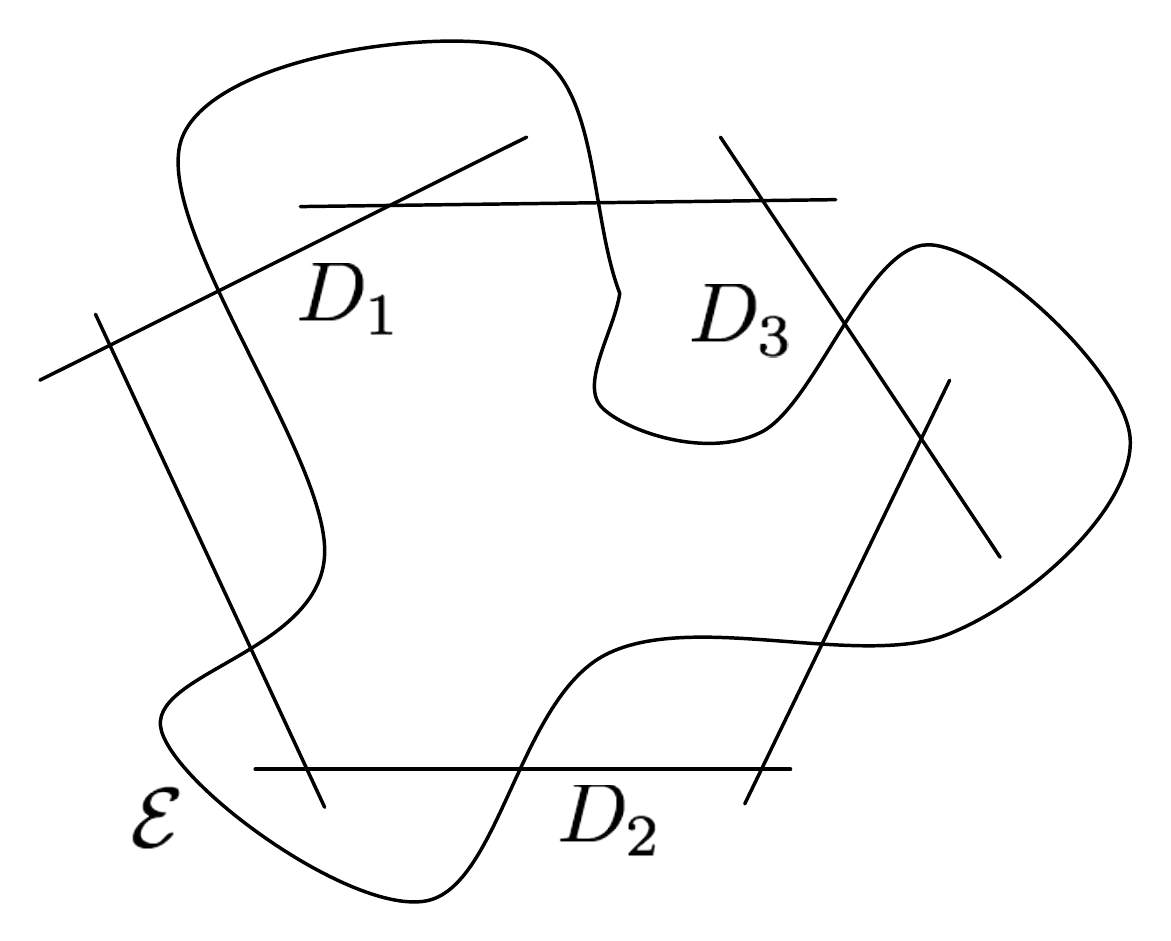}
  \caption{After blowup, the coordinate triangle becomes a hexagon in
    $P$ with three new divisors $D_i$. The elliptic curve $X_\circleddash=\{\cF_2(x,y;t)=0\}$ now meets each of the six divisors in one point.}
  \label{fig:blowup}
\end{figure}

The domain of integration for the sunset is the triangle
$\Delta=\{[x,y,z]\in \mathbb P^2| x,y,z\geq0\} $ and the second Symanzik
polynomial $\cF_2(x,y,z;t)=(x+y+z)(xy+xz+yz)-txyz$. The integral is given by
\begin{equation}
  I_3^2(t)= \int_{\Delta}  {z dx\wedge dy+x dy\wedge dz - y dx\wedge dz\over \cF_2(x,y,z;t)}\,.  
\end{equation}
This integral is very similar to the period integral in
equation~\eqref{e:periodsunset} for the elliptic curve
$\cEs:=\{\cF_2(x,y;t)=0\}$. The only difference between these two integrals is the domain of integration.  In the case of the period
integral in~\eqref{e:periodsunset} on integrates over a two-cycle and,
for well chosen values of $t$, the elliptic curve has no intersection
with the domain of integration, and therefore is a period of \emph{a pure
Hodge structure}. In the case of the Feynman integral
the domain of integration has a boundary, so it is not a cycle, and
for \emph{all} values of $t$ the elliptic curve intersects the domain
of integration. This is precisely because the
domain of integration of Feynman graph integral is given as
in~\eqref{e:DefDomain} that \emph{Feynman integrals lead to period of
mixed Hodge structures}.

As explained in
section~\ref{sec:motive} one needs to blow-up the points where the
elliptic curve $\cEs:=\{\cF_2(x,y,z;t)=0\}$ (the graph polar part) intersects the boundary
of the domain of integration
$\partial\Delta\cap \cEs=\{[1,0,0], [0,1,0], [0,0,1]\}$. The
blown-up domain is the hexagon $\mathfrak h$ in figure~\ref{fig:blowup}. 
 The associated mixed Hodge structure is
given by~\cite{Bloch:2013tra} for the relative cohomology $H^2(\mathcal P-\cEs, \mathfrak h -
    \cEs\cap \mathfrak h)$

 \minCDarrowwidth.1cm 
\begin{equation}\begin{CD} 
0 @>>> H^1(\mathfrak h - \cEs\cap
    \mathfrak h) @>>> H^2(\mathcal P-E, \mathfrak h -
    \cEs\cap \mathfrak h) @>>> H^2(\mathcal
    P-\cEs,\mathbb Q) @>>> 0
\end{CD}
\end{equation}
and for the domain of integration we have the dual sequence
\minCDarrowwidth.1cm 
\begin{equation}\begin{CD} 0 @>>> H_2(P-E) @>>> H_2(P-\cEs, \mathfrak h -
    \cEs\cap \mathfrak h) @>>> H_1(\mathfrak h - \cEs\cap \mathfrak h) @>>>
    0 
\end{CD}
\end{equation}

The Feynman integral for the sunset graph  coincides with  $I_3^2(t)=\langle
\omega,s(1)\rangle$ where $\omega$ in $F^1H^1(\cEs,\IC)$ is an element in the smallest Hodge filtration piece
$F^2H^1(\cEs,\IC)(-1)$, and $s(1)$ is a section in $H^1(\cEs,\IQ(2))$~\cite{Bloch:2013tra}.

The integral is expressed as the following combination of elliptic
dilogarithms 
\begin{equation}\label{e:Intsunset}
 -{I_3^2(t)\over 6}=-i{\pi\over6}\,\varpi_r(t)) (1-2\tau)
+ {\varpi_r(t)\over\pi} \,E_\circleddash(\tau) \,,
\end{equation}
where the Hauptmodul $t={\pi\over\sqrt3}\,
  \eta(\tau)^6\eta(2\tau)^{-3}\eta(3\tau)^{-2}\eta(6\tau)$, the real period $\varpi_r(t)={\pi\over\sqrt3}\,
  \eta(\tau)^6\eta(2\tau)^{-3}\eta(3\tau)^{-2}\eta(6\tau)$  and $\tau$
  is the period ratio for the elliptic curve $\cEs$.  Using
  $q:=\exp(2i\pi \tau)$ the   elliptic dilogarithm  is given by 
\begin{eqnarray}\label{e:Esunset}
    E_\circleddash(\tau) &=&- {1\over2i} \sum_{n\geq0}  \left(\Li2(q^n\zeta_6^5)
    +\Li2(q^n\zeta_6^4)-\Li2(q^n\zeta_6^2) -\Li2(q^n\zeta_6)\right)\cr
&+&{1\over 4i} \, \left(\Li2(\zeta_6^5)
    +\Li2(\zeta_6^4)-\Li2(\zeta_6^2) -\Li2(\zeta_6)\right)\,.
\end{eqnarray}
which we can write as well as $q$-expansion
\begin{equation}
  E_\circleddash(\tau)={1\over2}\,\sum_{k\in \mathbb Z\backslash \{0\}}{(-1)^{k-1}\over k^2}\, {\sin({n\pi\over3})+\sin({2n\pi\over3})\over
  1-q^k}\,.
\end{equation}
As we mentioned earlier this integral is not given by an elliptic
dilogarithm obtained by evaluating the real analytic function $D(z)$
to the contrary to the Mahler measure described in section~\ref{sec:mahler-measure}.

The amplitude is closely related to the {\it regulator} in arithmetic
algebraic geometry~\cite{Beilinson,BlochCMR,Soule,Brunault}. Let $conj: M_\IC \to M_\IC$ be the real
involution which is the identity on $M_\IR$ and satisfies
$conj(c\,m)=\bar c\,m$ for $c\in \IC$ and $m\in M_\IR$. With notation as
above, the extension class $s(1)-s_F\in H^1(\cEs,\IC)$ is well-defined
up to an element in $H^1(\cEs,\IQ(2))$ (i.e. the choice of $s(1)$). Since
$conj$ is the identity on $H^1(\cEs,\IQ(2))$, the projection onto the
minus eigenspace $(s(1)-s_F)^{conj=-1}$ is canonically defined. The
regulator is then 
\begin{equation}
\langle\omega,(s(1)-s_F)^{conj=-1}\rangle \in \IC\,.
\end{equation}

 \section*{Acknowledgements}
I would like to thank warmly Spencer Bloch for introducing me to the
fascinating world of mixed Hodge structure and motives. 
I would like to thank David Broadhurst for his comments on this text,
and Francis Brown for comments and
corrections,  as well for sharing insights on the relation
between quantum field theory amplitudes and periods. 
I would like to thank the organizers of string-math 2013 for the
 opportunity of presenting this work and writing this proceeding
 contributions.
PV gratefully acknowledges support from the Simons Center for Geometry and Physics, Stony Brook University at which some or all of the research for this paper was performed.
 This research of PV has   been supported by the ANR grant   reference QFT ANR 12 BS05 003  01, and the PICS 6076.

\bibliographystyle{amsalpha}

\end{document}